\documentclass[sigconf]{acmart}

\usepackage{booktabs}
\usepackage{multirow}
\usepackage{array}
\usepackage{xcolor}
\newcommand{\yes}{\textcolor{green!55!black}{\checkmark}}
\newcommand{\no}{\textcolor{red!70!black}{\ding{55}}}

\usepackage{pifont}

\setcopyright{none}
\renewcommand\footnotetextcopyrightpermission[1]{}
\begin{document}
\title{Scaling Author Identity Disambiguation to the World of Code: A Methodology}

\author{Audris Mockus}
\affiliation{\institution{University of Tennessee}\city{Knoxville}\country{USA}}

\begin{abstract}
We describe the methodology used to alias the free-text author/committer identities of
the entire World of Code (WoC) collection (version \texttt{V2604}, $\sim$107M distinct
author strings over $\sim$6B commits) into canonical persons, extending the
fingerprint-based anti-aliasing of ALFAA~\cite{amreen2019alfaa} and the 38M-id resolution
of Fry et al.~\cite{fry2020dataset} by an order of magnitude. The central engineering
problem at this scale is \emph{over-merge} rather than \emph{missed} merges: a small number of
bridge identities (bots, role accounts, placeholder emails, multi-author commit fields)
transitively weld otherwise-unrelated clusters through the global union step into
million-member ``mega-clusters.'' We report the full experimental record (twenty
experiments, including every unsuccessful approach) that led to the deployed design.
Node-level gates (information score, project spread, link degree) preserve recall but
cannot dissolve the mega; per-value blocklists of \emph{bad high-quality attributes} are
recall-safe but cannot break a redundant mesh; the working composition is a
betweenness-centrality cut over the exact union graph plus a per-edge classifier trained
on $2.6$M labels mined for free from GitHub no-reply ids. The same classifier, filtering
the expansion of dormant cross-project shingle groups, joined by GitHub's own account
assertions mined from no-reply ids, then recovers the recall the
precision work had foregone. Against human-adjudicated pairs the deployed map's per-edge
model transfers at AUC $0.99$; end to end, the largest cluster falls from $170{,}431$ (and
a predecessor's $3.0$M) to under $7{,}000$ while gold recall rises from $0.44$ to $0.70$
at \emph{increasing} precision, and on an independent $21$M-alias GitHub ground truth the
final map outscores both its predecessors and the published state of the art among
global, privacy-preserving resolvers. The record doubles as a catalog of scale lessons:
structural cuts do not transfer to edge sets they never saw, gradient boosting
shortcut-learns label-construction artifacts that a linear model survives, and
recall-only and precision-only benchmarks invert verdicts unless read together.
\end{abstract}

\maketitle

\section{Introduction}
\label{sec:intro}
Nearly every quantitative claim about open-source software rests on knowing \emph{who}:
developer productivity and turnover, project health and bus factors, the provenance chains
behind supply-chain security, the demographics of contribution. Git records authorship as
a free-text string, and developers scatter their work across name spellings, work and
personal emails, usernames, and, increasingly, deliberate anonymization. Resolving these
strings to persons (de-aliasing) is therefore foundational infrastructure for mining
software repositories, and getting it wrong is worse than not doing it: a resolver that
welds strangers together fabricates super-developers, corrupts network analyses, and
overrides explicit privacy choices.

This paper documents, as a complete experimental record, the construction of the author
identity map for World of Code (WoC)~\cite{ma2019woc} version \texttt{V2604}:
$\sim$$107$M distinct author ids over $\sim$$6$B commits, an order of magnitude beyond the
largest published resolutions~\cite{fry2020dataset}. At this scale the dominant failure
mode inverts from the literature's usual concern. Classical de-aliasing maximizes recall
against missed merges; in a global transitive union over $10^8$ identities, the binding
constraint is \emph{over-merge}: a handful of bridge identities (placeholder emails,
role accounts, template author strings, common-name homonyms) weld unrelated clusters
into mega-components that absorb millions of people. Our predecessor production map
carried a $3.0$M-member cluster; the first ungated union of \texttt{V2604} produced a
$170{,}431$-member one. Most of this paper is the systematic dissolution of that cluster
without sacrificing the legitimate merges around it, followed by the recovery, once
precision machinery existed to make it safe, of the recall that caution had deferred.

The paper makes five contributions. \emph{(1)~A methodology-as-record}: all twenty
experiments are reported, including the failures, because the failures carry the design
information: three families of node-level gates and an exhaustive attribute-blocklisting
arc had to fail, measurably and for articulable reasons, before the working design was
identifiable. \emph{(2)~A diagnosis}: over-merge at scale is driven by \emph{bad
high-quality attributes}, values rare enough to look identifying yet semantically void
(Cloudflare relay hashes, unconfigured git templates), and we separate them by intent
into \emph{privacy} masks (chosen; must never be merged) and \emph{homonym} defaults
(accidental; re-linkable by later evidence). \emph{(3)~Labels for free}: GitHub no-reply
ids embedded in raw commit strings yield $2.6$M labeled identity pairs at zero annotation
cost; a per-edge logistic classifier on gate signals turned features transfers to
human-adjudicated labels at AUC $0.99$, matching an actively-learned random forest that
required behavioral fingerprints~\cite{amreen2019alfaa}. \emph{(4)~A composed design}:
an exact reconstruction of the union graph, a sampled-betweenness cut of its $2{,}000$
load-bearing nodes, per-edge pruning of the residual homonym fragments, and
classifier-filtered expansion of dormant cross-project shingle groups, completed by
GitHub-asserted same-account edges, each mechanism
validated end-to-end, with the recall stages together raising gold recall from $0.44$ to
$0.70$ while precision \emph{rose}. \emph{(5)~External evaluation at adversarial scale}: on the $21$M-alias
single-author-repository ground truth of Bock et al.~\cite{bock2025dealiasing}, the final
map outperforms both predecessor maps and the published global state of the art, while
remaining mega-free. The benchmark's recall-only construction, read against the
precision-only gold audit, illustrates why identity resolvers must be graded on both axes
at once.

The map described here is deployed as the production identity layer of WoC
\texttt{V2604}. Section~\ref{sec:method} presents the pipeline, the experimental record
(Table~\ref{tab:explog}), and the analyses; Section~\ref{sec:concl} distills the lessons
we believe transfer beyond this system.

\section{Methodology}
\label{sec:method}

\subsection{Problem and Prior Work}
ALFAA~\cite{amreen2019alfaa} disambiguates developer identities by combining string
similarity with behavioral fingerprints under active learning; Fry et
al.~\cite{fry2020dataset} scaled a variant to 38M author ids over 2B commits. WoC
\texttt{V2604} contains roughly an order of magnitude more identities. At this scale the
precision failure mode inverts: a single high-connectivity identity that links to many
distinct people produces a transitive ``mega-cluster'' that absorbs millions of unrelated
authors. The methodology below is therefore organized around \emph{detecting and gating
bridge identities} rather than maximizing recall.

\subsection{Data and Pipeline}
The pipeline is two-phase, mirroring ALFAA's local-then-global structure:
\begin{description}
  \item[Inputs.] \texttt{CmtV2604.split} (per-id fingerprint fields:
    name $f$, family-name $l$, username $u$, email $e$, GitHub numeric id $g$);
    \texttt{P2aFull.V2604.\{0..31\}.s} (project$\to$author, project-sharded);
    \texttt{a2PFull.V2604.\{0..31\}.s} (author$\to$\emph{deforked} project $P$,
    author-sharded and author-sorted).
  \item[Phase 1 (local).] Within-project co-authorship/fingerprint linking
    (\texttt{aliasLink}) emits candidate same-person pairs. For \texttt{V2604} this
    yielded $49.76$M links over $70.3$M raw clusters.
  \item[Phase 2 (global).] Union--find over the pairs, then per-cluster representative
    selection (\texttt{selectRep}). The representative maximizes an information score
    \(\mathrm{info}(id)=\sum_{x\in\{f,l,u,e,g\}}\log\!\frac{N_x}{\mathrm{freq}_x(\cdot)}\),
    i.e.\ rarer field values are more identifying. ``Bad'' ids (bots/generic/machine-local,
    from \texttt{getUninformative}) get $-\infty$ and are never representatives.
\end{description}

\subsection{The Over-Merge Phenomenon}
Run unconditionally, Phase~2's union step welds clusters through bridge identities. The
largest resulting cluster contained $111{,}926$ members spanning $857{,}853$ distinct
projects, clearly not one person. Manual triage attributed the bridging to three identity
types: multi-name commit fields (``A and B''), shared role/service accounts, and
rare-but-shared emails. The remaining experiments test mechanisms for suppressing these
bridges without discarding legitimate small-cluster merges.

\subsection{Gating Strategies}
We define an identity as \emph{eligible to bridge} only if it passes a gate; ineligible ids
still alias to themselves but cannot transitively union others. Two gate signals were
evaluated.

\paragraph{(A) Information-score cutoff.}
\texttt{selectRepGated} unions a pair only if both endpoints are non-bad, non-multi-name,
and have $\mathrm{info}>\textsc{cutoff}$. The hypothesis: uninformative ids are the bridges.

\paragraph{(B) Project-spread bot-gate (``opposite approach'').}
\texttt{selectRepSpread} gates on \emph{deforked} project spread $n_P(id)$ computed from
\texttt{a2PFull} (uppercase $P$ = fork-collapsed; lowercase $p$ would over-count because
forks share git history). A pair unions only if both endpoints are non-bad, non-multi-name,
and $n_P<\textsc{T}$. The hypothesis: bridges are exactly the ids present in implausibly many
\emph{unrelated} projects (bots, role accounts).

\paragraph{(C) Link-graph degree-gate.}
A third signal targets the bridges that (A) and (B) both miss. We compute each identity's
\emph{degree} $d(id)$, the number of distinct Phase-1 partners it links to, directly from the
link list, and gate a pair only if both endpoints are non-bad, non-multi-name, and $d<\textsc{D}$.
The hypothesis: the identities that hold the residual mega-cluster together are not prolific in
\emph{projects} but prolific in \emph{links}: shared/template author strings that co-occur with
thousands of distinct real people inside a single dense corpus. Because the loader of (B) already
gates on a generic ``\texttt{value;id}'' file, the same code serves the degree-gate by swapping
the spread file for a degree file.

\begin{table*}[t]
\centering
\caption{Experiment log for over-merge suppression in WoC \texttt{V2604} aliasing.
Baseline = unconditional global union. ``size-2'' is the count of legitimate two-member
clusters (a recall proxy we want to preserve); ``top'' is the largest cluster size
(an over-merge proxy we want small).}
\label{tab:explog}
\footnotesize
\begin{tabular}{@{}clp{4.6cm}p{4.8cm}c@{}}
\toprule
\# & Approach & Hypothesis / setup & Outcome & St. \\
\midrule
1 & Ungated union (baseline) &
  Global union over all $49.76$M Phase-1 pairs; rep = max info-score. &
  Over-merge: top cluster $=111{,}926$ members across $857{,}853$ projects;
  $70.3$M clusters total. Bridges = multi-name fields, role accounts, shared emails. &
  \no \\
\addlinespace
2 & Info-score cutoff gate &
  Union only if both endpoints non-bad, non-multi-name, $\mathrm{info}>\textsc{cut}$.
  Swept $\textsc{cut}\in\{16,24,32,48,56,60,64,96\}$ (Table~\ref{tab:infocut}). &
  Blunt. Union rate flat ($\sim$93\%) then cliff $48\!\to\!64$. Only $\textsc{cut}=64$
  dissolves mega ($>$10k $\to0$) but collapses size-2 from $9.60$M$\to2.14$M. Cannot
  separate low-info from sparse-genuine. &
  \no \\
\addlinespace
3 & Project-spread bot-gate &
  Union only if both endpoints non-bad, non-multi-name, deforked spread $n_P<\textsc{T}$.
  $n_P$ from \texttt{a2PFull.V2604} (fork-collapsed). Swept
  $\textsc{T}\in\{200,500,1000,2000,5000\}$ (Table~\ref{tab:spread}). &
  Recall pristine (size-2 fixed at $9.60$M for all \textsc{T}) but the mega dissolves only
  slowly with more gating (top $51{,}860\!\to\!28{,}526$ from \textsc{T}$=200\!\to\!100$); each
  halving of the giant costs $\sim$4$\times$ more gated ids. The residual mega is welded by a
  long tail of \emph{low}-spread bridges no spread threshold reaches. Beats Exp.~2 on recall but
  cannot dissolve the mega alone. &
  \no \\
\addlinespace
4 & Link-graph degree-gate &
  Gate on co-link \emph{degree} $d(id)$ (distinct Phase-1 partners), reusing the Exp.~3 loader
  with a degree gate file. Hypothesis: the low-spread residual bridges are high-\emph{degree}
  shared/template author strings. &
  Signal strongly validated: degree is far more concentrated than spread (only $296$ ids with
  $d\ge500$, $54$ with $d\ge1000$; max $4380$) and its top nodes are precisely the residual-mega
  bridges: French legal-text template strings, bare \texttt{=}/\texttt{<=>}, machine-local
  empties (Table~\ref{tab:degree}), all \emph{low}-spread, so invisible to Exp.~3. But the gated
  sweep slow-peels exactly like Exp.~3 (top $71{,}172\!\to\!20{,}635$ for D$=1000\!\to\!50$; $>$10k
  never $<2$): each threshold gates the top bridge and exposes the next. Even the
  \emph{combined} union-gate $d{\ge}100\lor n_P{\ge}100$ ($42$k ids, recall pristine) only reaches
  one residual mega (top $25{,}957$), welded by a moderate-degree/spread Cloudflare hash string no
  node threshold reaches (Table~\ref{tab:deggate}). No node-gate dissolves the mega cleanly. &
  \no \\
\bottomrule
\end{tabular}
\end{table*}

\begin{table*}[t]
\centering
\caption{Experiment log for over-merge suppression in WoC \texttt{V2604} aliasing (continued).}
\footnotesize
\begin{tabular}{@{}clp{4.6cm}p{4.8cm}c@{}}
\toprule
\# & Approach & Hypothesis / setup & Outcome & St. \\
\midrule
5 & Cross-ghid false-link rate (edge labels) &
  Free ground truth: a Phase-1 link whose endpoints carry agreeing GitHub numeric ids is a true
  positive, conflicting ids a false link. Mine the id from the raw \texttt{NNNN+login@\dots
  noreply.github.com} prefix; cross-tabulate the false rate by endpoint degree
  (Table~\ref{tab:ghidrate}). &
  $29.95$M ids labeled; $2.64$M links labeled on both ends; \textbf{$11.4\%$ are false}. The rate
  rises sharply with degree ($5.7\%$ at $d{<}10$ to \textbf{$99.8\%$ at $d{\in}[100,499]$}), an
  \emph{independent} confirmation that high-degree ids are bridges (Exp.~4), and the label set for
  a per-edge classifier. &
  \yes \\
\addlinespace
6 & Learned per-edge classifier &
  Replace node-gating with a pairwise model on the $2.64$M ghid-labeled links (Exp.~5). Features =
  the gate signals turned per-edge: endpoint degree (min/max), spread (min/max), info-score
  (min/max), component-match flags (f/l/u/e/domain), and \emph{matchScore} ($\sum\log(N/\mathrm{freq})$
  over matching components). Logistic regression, $300$k train / $2.34$M test
  (Table~\ref{tab:edgeclf}). &
  Test accuracy $95.2\%$ vs.\ a $88.6\%$ predict-all-true baseline; \textbf{AUC $=0.956$}. The
  false rate is monotone in both degree (the over-merge axis) and matchScore (the true-link axis);
  the model keeps a high-degree id's strongly-matched links and drops only its weak high-degree
  welds, exactly what no node-gate could do. \emph{External validation} on the human-labeled ALFAA
  gold set~\cite{amreen2019alfaa} ($469$k pairs, all $2{,}345$ ids found in \texttt{V2604}):
  in-domain 5-fold AUC $=0.994$; the ghid-trained model \emph{transfers} to the human labels at AUC
  $=0.987$, matching ALFAA's random forest \emph{without} its doc2vec behavioral fingerprint. &
  \yes \\
\addlinespace
7 & Flat edge-probability threshold &
  Apply the Exp.~6 classifier to score all $49.76$M Phase-1 links, keep those with $p\ge\tau$, and
  re-run the global union. Sweep $\tau\in\{0.30,0.50,0.70,0.90\}$ and check whether any $\tau$
  dissolves the residual mega ($>$10k$\,\to0$) while holding size-2 near baseline
  (Table~\ref{tab:tausweep}). &
  The mega survives $\tau{=}0.30$ and $0.50$ ($>$10k$\,=1$) and dies only at $\tau{\ge}0.70$, but
  by then size-2 has fallen to $7.78$M and $2.6$M ids have dropped to singletons. The trade is
  \emph{monotone}: no $\tau$ both kills the mega and preserves recall, because the mega's welds
  share an \emph{exact} placeholder email and so the classifier scores them \emph{high}
  ($e/\mathrm{dom}/\mathrm{rawJac}$ all fire). A global threshold cannot separate them. &
  \no \\
\addlinespace
8 & Bad high-quality attribute detection &
  Root-cause the high-scoring welds: an attribute value is a \emph{bad bridge} if a single exact
  string ties together many \emph{distinct} person-names. Stream \texttt{Cmt.split.e} (sorted by
  email) and count distinct $\langle f,l\rangle$ names per email; classify each by \emph{intent}
  (Table~\ref{tab:badattr}). &
  $3.84$M emails bridge $\ge3$ distinct names. A recall-safe two-tier rule (lexical placeholders at
  $\ge3$; form-only signals such as github-noreply, long-numeric QQ, and pure spread only at $\ge10$) yields
  \textbf{$91{,}432$} bad emails ($90{,}011$ new beyond the $4{,}776$-entry manual stoplist),
  split by intent into \emph{privacy} (deliberate anonymisation: \texttt{123@}, \texttt{anonymous},
  \texttt{johndoe@}, Cloudflare/github-noreply hashes) vs.\ \emph{homonym} (unintentional defaults:
  \texttt{root}, \texttt{admin}, \texttt{*.local}). Neutralising these at source removes the welds
  \emph{before} scoring, the fix Exp.~7 shows a threshold cannot make. &
  \yes \\
\bottomrule
\end{tabular}
\end{table*}

\begin{table*}[t]
\centering
\caption{Experiment log for over-merge suppression in WoC \texttt{V2604} aliasing (continued).}
\footnotesize
\begin{tabular}{@{}clp{4.6cm}p{4.8cm}c@{}}
\toprule
\# & Approach & Hypothesis / setup & Outcome & St. \\
\midrule
9 & Source-blank emails + re-union &
  End-to-end test: blank all $91{,}432$ bad emails in the commit table, re-run Rule~2, recompute the
  global union (Table~\ref{tab:blankreunion}). Does removing bad-email welds at source dissolve the
  mega while preserving recall? &
  \emph{Recall-safe but insufficient.} Blanking voids $1.07$M fields and drops $403{,}635$
  email-only welds; size-2 \emph{rises} $9{,}304{,}193\to9{,}307{,}723$ (vs.\ Exp.~7's collapse):
  a source-neutralised value severs only spurious welds. But the mega persists ($>$10k $=1$): it is
  \emph{multiply connected} through placeholder names/usernames too. Bad-attribute neutralisation is
  necessary and recall-safe, but must cover \emph{every} attribute (or pair with a structural cut). &
  \no \\
\addlinespace
10 & Multi-attribute source-blanking &
  Extend Exp.~8's spread detector to \emph{usernames} (distinct names per login) and \emph{names}
  (distinct emails per $\langle f,l\rangle$), classify by the same intent taxonomy, and blank all
  three blocklists ($91{,}432$ emails $+$ $12{,}418$ logins $+$ $628$ names; $5.46$M fields over
  $106.8$M rows) before re-running \emph{all three} Phase-1 link generators and the union. &
  \emph{Recall-safe but still insufficient.} Clusters $+112$k, $117$k ids drop to singletons,
  size-2 unchanged ($9{,}304{,}193\to9{,}304{,}250$); but the mega shrinks only $6.4\%$
  ($170{,}431\to159{,}483$, same representative). Side finding: the $+17.6$M ``pairs'' in the
  treatment log is an artifact: shingle lines parse as \texttt{type;anchor} pseudo-node unions
  invisible to the member stats, so the union is driven by Rules~1--2 alone ($84$M links).
  The mega is welded by a redundant mesh no per-value blocklist reaches. &
  \no \\
\addlinespace
11 & Structural bridge detection &
  Reconstruct the production union graph exactly in NetworKit ($45.1$M eligible ids, $41.7$M
  unique Rule~1--2 links; giant $=170{,}431$, matching the union mega to the node). On the mega
  subgraph ($347$k edges, mean degree $4.1$) compute sampled betweenness centrality
  ($4{,}000$-source Brandes), articulation points (iterative Tarjan), and $k$-cores;
  rank candidate cut-nodes and test overlap with the attribute blocklists. &
  \textbf{Betweenness finds what no value rule can.} Two bridge populations emerge: uncaught
  placeholder \emph{hubs} (\texttt{=\,<=>} deg $3{,}216$, betw.\ $0.258$; Gerrit service/relay
  accounts; Badger bots) and, invisible to every attribute detector, \emph{homonym chains} of
  moderate-degree common-surname ids (kim/lee/smith/unknown dominate the top-$2$k last-tokens).
  Of the top-$2{,}000$ bridges, $0$ are in \texttt{bad.ids} and only $32$ carry a blocklisted
  email ($98.4\%$ attribute-clean). Cut sweep (Table~\ref{tab:cutsweep}): removing the top
  $2{,}000$ by betweenness ($1.2\%$ of mega nodes) shatters the mega to a largest component of
  $7{,}268$; top-$5{,}000$ $\to1{,}813$ with $89.5\%$ of survivors freed into $\le$100-size
  components. $28\%$ of mega nodes are articulation points. \textbf{End-to-end: gating these
  $2{,}000$ ids in the production union dissolves the mega} ($>$10k $1\to0$; top
  $170{,}431\to7{,}268$, exactly the local prediction) \emph{with recall above baseline}
  (size-2 $9{,}304{,}193\to9{,}305{,}460$). The first design in the record to achieve both. &
  \yes \\
\addlinespace
12 & Classifier refinement of residual fragments &
  Extract the ten residual mega fragments of size $\ge1{,}000$ left by the $K{=}2{,}000$ cut
  ($29{,}315$ ids, $61{,}637$ internal edges, exactly the new $1$k--$10$k clusters of the gated
  union), score every internal edge with the Exp.~6 classifier, and prune at $p<\tau$ \emph{inside
  the fragments only} (unscored edges kept). &
  \textbf{Targeted pruning shatters the homonym blocks a global threshold could not afford to.}
  Member samples confirm the fragments are \emph{given-name} homonym blocks (David, Michael,
  Daniel, Thomas, James, Jan, Chris), a Korean-surname block (Kim), an institutional-domain block
  (RIT students), and a dense Gerrit-relay remnant (Table~\ref{tab:fragclf}). At $\tau{=}0.5$ the
  largest surviving sub-component is $1{,}942$ (from $7{,}268$) and $\sim$$46\%$ of fragment ids
  are freed into components $\le$100, at a recall risk bounded to $0.03\%$ of all ids, via edges
  the model itself certifies as unlikely. Contrast Exp.~7: the \emph{same} classifier applied
  globally cost $2.6$M singletons. Caveats: the Kim and RIT blocks resist (freed $27$/$26\%$;
  shared exact emails and \texttt{@rit.edu} domains score high), and the Gerrit fragment is only
  $16\%$ scored (relay ids lack parseable features). &
  \yes \\
\bottomrule
\end{tabular}
\end{table*}

\begin{table*}[t]
\centering
\caption{Experiment log for over-merge suppression in WoC \texttt{V2604} aliasing (continued).}
\footnotesize
\begin{tabular}{@{}clp{4.6cm}p{4.8cm}c@{}}
\toprule
\# & Approach & Hypothesis / setup & Outcome & St. \\
\midrule
13 & ALFAA gold pairs as arbiter (and audit of the gold itself) &
  Score V3 and production V2604 against the $469{,}369$ human-rated ALFAA pairs, of which $264{,}346$
  are unique non-self pairs over $2{,}345$ ids with $1{,}644$ true matches (the $2{,}345$ trivial
  self-pairs excluded): a map predicts ``same'' iff both ids share a representative
  (Table~\ref{tab:goldarbiter}); decompose every disagreement and lexically audit the labels. &
  \textbf{The mega is V3's recall engine \emph{and} its precision sink; V2604's misses are
  genuine, structural, and largely recoverable.} V3: recall $1.000$, precision $0.522$; $180$
  gold ids ($7.7\%$) sit inside its $3.0$M mega, where $94.6\%$ of its $1{,}505$ FPs are
  manufactured while buying only $139$ TPs ($10{:}1$); mega-free V3 scores $0.915/0.949$.
  V2604: precision $0.873$ ($104$ FPs), recall $0.436$; $95.8\%$ of misses are alias/alias
  (not gate-imposed), incl.\ exact-same-email pairs; cause: Phase-1 requires project
  co-occurrence and Rule-3 shingles are inert (Exp.~10), yet $83.9\%$ of missed pairs already
  sit in a computed shingle group ($87.3\%$ of all matches $\Rightarrow$ recall ceiling
  $\sim$$0.87$ via classifier-filtered expansion; naive expansion unsafe: one username group
  spans $23$ distinct people). Gold audit: $20$ of V2604's $104$ ``false'' positives share an
  exact email/name/handle the raters overrode; $\sim$$6$ of $9$ zero-overlap ``matches'' look
  mislabeled, while the $3$ real ones (handles \texttt{paulczar}, \texttt{dimtruck}) V2604
  co-clusters without any mega. &
  \yes \\
\addlinespace
14 & Classifier-filtered shingle expansion &
  Claim the recall ceiling Exp.~13 located: expand every Rule-3 shingle group into member
  pairs, score each with the Exp.~6 classifier, and admit pairs with $p\ge\tau$ as
  \emph{first-class union edges} alongside the production inputs (betweenness gate $+$
  fragment pruning). $66{,}459{,}771$ unique expansion edges score $p\ge0.5$; sweep
  $\tau\in\{0.5,0.7,0.9\}$ end-to-end (Table~\ref{tab:shingletau}). &
  \textbf{$\tau{=}0.9$ adds $30.8$M cross-project edges with a clean profile.} At
  $\tau{=}0.5$ the mega \emph{returns} ($2{,}292{,}596$ members): the betweenness cut was
  computed on the \emph{pre-expansion} graph, so the new edges re-weld around the gated
  bridges; a structural cut does not transfer to an edge set it never saw. $\tau{=}0.7$
  still leaves two $>$10k clusters. At $\tau{=}0.9$ ($46\%$ of scored edges): $>$10k $=0$,
  top $=3{,}862$, unioned ids $110.6$M$\,\to141.4$M ($+30.8$M), size-2
  $9{,}307{,}433\to10{,}189{,}048$ ($+881{,}615$), the largest recall gain of any
  experiment, at the cost of $14$ clusters in the $1$k--$10$k band (vs.\ $4$ in
  production). \textbf{Gold verdict: recall $0.4361\to0.6454$ with precision
  \emph{rising} $0.8733\to0.8820$} (TP $717\to1{,}061$, FP $104\to142$), the first
  experiment to move recall materially, and it escapes the Exp.~7 monotone trade
  entirely. &
  \yes \\
\addlinespace
15 & Enriched per-pair classifier (v2) &
  Re-arm the scorer before re-scoring the expansion: add true deforked-project IoU
  (exact $n_P$ $+$ capped $100$-hash project sets from \texttt{a2PFull}, fixing the
  \texttt{spread20} truncation that zeroed $n_P$ for $99.4\%$ of ids), $\log(1{+}x)$ on
  heavy-tailed counts, per-value \emph{spread} of each matched attribute, two
  name$\leftrightarrow$email-local-part features (\texttt{f.l}-pattern and token-in-handle),
  and a Wiktionary given-name flag; compare logistic vs.\ gradient boosting
  (Table~\ref{tab:edgeclf2}). &
  \textbf{Logistic v2.1 reaches gold-transfer AUC $0.9910$} (v1: $0.9861$). LGBM wins
  in-distribution (test AUC $0.9990$) but \emph{collapses on gold to $0.56$}: shortcut
  learning: $74\%$ of its split gain is on \texttt{rawJac3}, which leaks the shared
  \texttt{noreply} handle inside both email strings of every ghid-labeled pair; trees
  threshold the leak, the linear model uses it benignly. Ablating \texttt{rawJac3} recovers
  LGBM to $0.981$, still below logistic. Caveats from the learned weights: project IoU is
  \emph{inert in-distribution} (ghid pairs nearly all co-occur; its value is on
  cross-project pairs), and matched-value spread enters \emph{positive} (synonym-richness
  confound), so spread cannot serve as a per-pair homonym signal without
  out-of-distribution labels. \textbf{Map-level verdict: v2.1 does not transfer.} Re-scoring
  all $256.7$M expansion pairs yields $54.0$M unique pairs $p\ge0.5$ (v1: $66.5$M;
  $\tau{=}0.9$: $26.1$M vs.\ $30.8$M), more selective, structurally cleaner
  ($1$k--$10$k $=9$ vs.\ $14$, $>$10k $=0$). The S2 map edges gold \emph{recall}
  ($0.6490$ vs.\ $0.6454$; precision $0.8789$ vs.\ $0.8820$) but \emph{loses} the
  GitHub benchmark outright: max9 recall $0.5118$ vs.\ $0.5886$, FullyMerged $0.4658$
  vs.\ $0.5403$; the pairs v2.1 prunes are exactly the within-handle mass Exp.~16
  counts. A better in-distribution AUC ($0.9910$ vs.\ $0.9861$) bought a worse map:
  \textbf{v1-scored S retained as the production candidate.} &
  \no \\
\bottomrule
\end{tabular}
\end{table*}

\begin{table*}[t]
\centering
\caption{Experiment log for over-merge suppression in WoC \texttt{V2604} aliasing (continued).}
\footnotesize
\begin{tabular}{@{}clp{4.6cm}p{4.8cm}c@{}}
\toprule
\# & Approach & Hypothesis / setup & Outcome & St. \\
\midrule
16 & GitHub-scale external recall benchmark &
  Evaluate the maps on the single-author-repository ground truth of Bock et
  al.~\cite{bock2025dealiasing}: $9{,}568{,}385$ GitHub handles owning $21{,}477{,}370$
  commit ids in their own single-author repos, $10^4\times$ larger than ALFAA,
  recall-only by construction (only within-handle pairs are visible; lumping is
  invisible). Their metrics: per-handle pairwise recall (macro-averaged), FullyMerged\%,
  ExclSingle\%, at handle-size caps max3/max9/all (Table~\ref{tab:ghgt}). &
  At max9: V3 recall $0.487$ / FullyMerged $0.442$; production V2604 $0.433$ / $0.395$;
  published baselines: WoC V2409 $0.450$ / $0.411$, GitAuthority $0.771$ / $0.736$. V2604
  sits \emph{below} its own predecessor on a recall-only yardstick: the over-merge
  surgery of Exps.~1--12 traded exactly the pair mass this benchmark counts, and V3's edge
  is partly mega-bought (its map also resolves only $19.2$M of the $21.5$M aliases vs.\
  V2604's $21.5$M). Confirms Exp.~13 externally at $10^4\times$ the scale. \textbf{The
  Exp.~14 expansion map then retakes the lead: recall $0.5886$ / FullyMerged $0.5403$ at
  max9}, above V3, above published V2409, above GitAuthority's privacy mode ($0.478$),
  mega-free; remaining gap to full GitAuthority ($0.771$) = within-repo context $+$
  noreply mining. The v2.1-rescored map (S2) drops to $0.5118$ / $0.4658$ at max9
  despite its better gold AUC: this benchmark, not the in-distribution score, made
  the production call (S over S2). &
  \yes \\
\addlinespace
17 & Production alternatives: $\tau$ relaxation, ghid edges, unions; coverage metric &
  Three user directives reframe the selection: (a) explore all promising edge sources;
  (b) treat both benchmarks as noisy (gold has mislabels; large same-ghid groups are
  shared/org accounts, not persons); (c) judge maps also by \emph{commit-weighted
  coverage}, counting low-quality ids matchable within a project. Candidates beyond S/S2:
  v2.1 at $\tau{=}0.8$ ($34.1$M edges); $5{,}736{,}921$ same-ghid pairs from
  \texttt{NNNN+login} noreply ids (groups of $2$--$20$ only; $346$ larger groups up to
  $12{,}140$ ids excluded as shared accounts); and unions thereof. Every new edge set
  passes a STATS mega-check before any map (Exp.~14 lesson). &
  All STATS-clean ($>$10k $=0$). S2@$\tau$0.8: gold recall $0.7019$ /
  precision $0.8789$ (TP $1{,}154$), GH max9 recall $0.5828$ / FM $0.5354$. S$+$ghid:
  gold \emph{identical} to S ($0.6454/0.8820$: the gold set has almost no noreply ids)
  but GH max9 recall $0.5995$ / FM $0.5516$: GitHub's own account assertion is
  orthogonal evidence. The two gains are independent and \emph{compose}: the triple
  union SUG (v1@$0.9\,\cup$ v2.1@$0.8\,\cup$ ghid; $181{,}156{,}655$ unioned pairs,
  largest $6{,}910$, a single noreply identity) beats the pairwise SG8 on \emph{both}
  benchmarks: gold $\mathbf{0.7026}/\mathbf{0.8790}$ (TP $1{,}155$, FP $159$; SG8
  $0.7019/0.8789$) and GH max9 recall $\mathbf{0.6035}$ / FM $\mathbf{0.5558}$ (SG8
  $0.5931/0.5460$), and tops commit-weighted coverage: $73.5\%$ of all
  $5.87$B commits land in a multi-id cluster (B $66.3\%$, S $72.7\%$, SG8 $73.2\%$).
  \texttt{a2AFullSUG.V2604.s.gz} is the production map, superseding
  \texttt{a2AFullB}. &
  \yes \\
\bottomrule
\end{tabular}
\end{table*}

\begin{table*}[t]
\centering
\caption{Experiment log for over-merge suppression in WoC \texttt{V2604} aliasing (continued).}
\footnotesize
\begin{tabular}{@{}clp{4.6cm}p{4.8cm}c@{}}
\toprule
\# & Approach & Hypothesis / setup & Outcome & St. \\
\midrule
18 & Within-project matching of low-quality ids; commit-coverage gain &
  The production map deliberately self-maps $5.77$M bad/local ids (generic names,
  machine-local addresses) because they are ambiguous \emph{across} projects; within a
  single deforked project $P$ their reuse is usually unambiguous. Matcher rules:
  \emph{anchor}: the bad id's exact lowercased full name or email local-part matches
  the ids of exactly one person in $P$ (a $\sim$50-token stoplist of generic names and
  a 3-character minimum guard the anchor); \emph{solo}: $P$ has exactly one person, so
  unmatched bad ids default to them; anything ambiguous stays self-mapped, bots are
  dropped. Primary metric (user directive): \emph{commit coverage of human IDs}, i.e.\ by
  how much does within-project expansion raise the share of commits attributed to a
  human identity over the cross-project map alone? Measured directly at commit level
  via provenance tags (global $g$ $|$ bot $B$ $|$ within-project $p$ $|$ self $s$) in
  the new \texttt{c2AFull} table over all $5{,}866{,}595{,}698$ commits. The bad/local
  pool that survives the cross-project map and the bot filter is $p+s =
  195{,}352{,}712$ commits ($3.33\%$ of all; a commit-weighted count that supersedes
  the earlier $0.61\%$ distinct-id estimate). &
  \textbf{Done.} Phase 1: $189{,}359{,}993$ projects scanned, $11{,}352{,}698$
  bad-id occurrences, $2{,}355{,}347$ matched by anchor $+$ $3{,}003{,}834$ by solo
  $=$ $5{,}359{,}181$ recovered ($47.2\%$ of occurrences;
  \texttt{P2aAFull.V2604.s}: $P$;raw;$A$;rule), parallelized $32\times$ by
  sHash($P$) bucketing ($\sim$24h single-core $\to$ $\sim$45min/bucket).
  Phase 2 (provenance-tagged \texttt{c2AFull}, 128 shards): within-project resolved
  $p = 93{,}381{,}683$ commits $= 47.8\%$ of the $195.35$M bad/local pool ($141$
  conflicts, negligible). Commit coverage of \emph{human} (non-bot) IDs rises from
  $g/(g{+}p{+}s) = 96.49\%$ (cross-project map alone) to $(g{+}p)/(g{+}p{+}s) =
  98.17\%$, i.e.\ $+1.68$ pp; residual self/unresolved $s = 101{,}971{,}029$
  ($1.83\%$). Deliverable \texttt{A2clsFull.V2604.s} (\texttt{member;canonical;class},
  $106{,}826{,}059$ ids): good $94.37\%$ / bad-attr $2.48\%$ / local $2.40\%$ /
  bot $0.52\%$ / partial $0.23\%$ (partial $=$ only one of name/email), companion
  to the within-project layer \texttt{P2aAFull}. &
  \yes \\
\bottomrule
\end{tabular}
\end{table*}

\begin{table*}[t]
\centering
\caption{Experiment log for over-merge suppression in WoC \texttt{V2604} aliasing (continued).}
\footnotesize
\begin{tabular}{@{}clp{4.6cm}p{4.8cm}c@{}}
\toprule
\# & Approach & Hypothesis / setup & Outcome & St. \\
\midrule
19 & Cryptographic signatures as an identity anchor &
  A signed commit binds the free-text author string to a key the committer controls,
  so a shared signing key could supply \emph{verified} alias edges independent of the
  ghid (Exp.~5) and ALFAA (Exp.~13) evidence, and signing could serve as a positive
  trust signal for the bad-id detector (Exp.~8). Every signature already sits, unparsed,
  in the \texttt{gpgsig} block that the commit-table generator appends to the message
  field, so extraction is an isaac scan over \texttt{c2datFull} ($128$ shards, no
  object-database pass) yielding \texttt{c2sigFull} (\texttt{commit;sigtype}). Measure
  the corpus prevalence, the family split, and, the question that decides whether
  signatures can move the map, the id-class composition of signers: does signing reach
  the bad/local tail that Phase-1 cannot resolve (Table~\ref{tab:sigclass})? &
  \textbf{A precision and trust anchor, not a tail-recall fix.}
  $1{,}031{,}721{,}316$ of the $5{,}866{,}595{,}698$ commits are signed ($17.59\%$,
  exact over all $128$ shards), PGP $98.96\%$ / SSH $1.02\%$ / X.509-sigstore $0.02\%$.
  But signing concentrates in the population Rules~1--2 already merge: on a representative
  commit-hash shard ($1/128$), of $3{,}775{,}298$ distinct signing ids $99.52\%$ are
  good-class, while the hard-to-alias tail is far under-represented against the
  \texttt{A2clsFull} baseline (bad-attr $0.36\%$ vs.\ $2.48\%$; \emph{local}
  $0.004\%$ vs.\ $2.40\%$, a $500\times$ gap; overall $\sim$$14\times$ cleaner). So a
  signature verifies identities that are already resolvable and rarely touches the
  \texttt{root@localhost}/generic-name welds; its value to aliasing is an independent
  evaluation instrument and a per-id trust tier, not recall on the tail. The shared-key
  fanout gate (person vs.\ org/CI keys, seeded by the web-flow key
  \texttt{4AEE18F83AFDEB23}) mirrors the Exp.~8 name-spread gate and is the next step. &
  \yes \\
\midrule
20 & Key fan-out and the attestation gate &
  A raw signed rate of $17.59\%$ (Exp.~19) counts every commit whose message carries a
  \texttt{gpgsig} block, but many of those keys are shared platform or automation keys that
  bind to thousands of authors and would over-merge if trusted as alias edges. Parse the
  OpenPGP signature packet to its issuer (subpacket 33 fingerprint, else subpacket 16
  key-id, normalized to \texttt{keyid16}), build the key$\leftrightarrow$author bipartite
  graph, and apply a fan-out gate ($>$$50$ distinct canonical authors $\Rightarrow$ shared,
  the crypto analogue of Exp.~8): does a per-key attestation isolate a near-injective
  personal-key population, and do any of those personal keys span two or more canonical ids
  that the map split apart (a recall-repair edge)? &
  \textbf{Shared keys dominate the raw rate; the gate isolates a large injective personal
  layer.} \texttt{pgpissuer.pl} (gpg-validated) resolves an issuer for $99.92\%$ of parsed
  PGP signatures. Corpus-wide, $39{,}077{,}350$ signed authors associate with $586{,}011$
  keys ($44{,}871{,}077$ author--key pairs); the two heaviest keys carry $95.9\%$ of those
  associations: the GitHub web-flow key \texttt{4AEE18F83AFDEB23} alone spans $30{,}440{,}944$
  authors and one second platform key $12{,}592{,}731$. The fan-out gate ($>$$50$) drops the
  $2{,}651$ shared keys ($0.45\%$, carrying $97.6\%$ of associations), leaving $583{,}360$
  personal keys, $72.20\%$ ($423{,}095$) bound to a single author. The aliasing payoff is
  broad: $156{,}397$ personal keys (fan-out $2$--$20$) link $549{,}388$ raw author strings,
  i.e.\ $392{,}991$ candidate same-person associations that Exp.~21 calibrates against the map. &
  \yes \\
21 & Signature-attested alias gold + map calibration &
  Treat each person-key's co-signed raw strings as same-person gold; map them through
  \texttt{a2AFullSUG} and measure recall (co-signed pairs the map already merges), triaging
  misses into genuine recall gaps vs.\ shared-key false positives. A cryptographic gold larger
  than ALFAA and free of hand labeling. &
  \textbf{Recall is real but partial, and signatures alone are not a clean gold.} On the clean
  fan-out-$2$ tier ($86{,}929$ keys) the map already unifies $63.0\%$; recall falls monotonically
  with fan-out ($0.63\!\to\!0.01$ over $2\!\to\!21$--$50$), the shared-key confound made visible.
  Of the $32{,}192$ fan-out-$2$ splits, only $9{,}036$ ($28.1\%$) carry a corroborating attribute
  (shared localpart / non-generic domain / name token) = genuine recall-repair edges; the other
  $71.9\%$ are two people on one key, correctly unmerged. Signatures need attribute corroboration
  to serve as same-person gold; corroborated, they yield $9{,}036$ high-precision repair edges. &
  \yes \\
22 & Impersonation / vanity-id detection via key-dispersion &
  Invert the fan-out gate: per author \emph{string}, count distinct signing keys. One keyholder
  maps to few keys; a string signed by many keys is shared, mirrored, or impersonated. A
  signature-derived alternative to the hardcoded bad-author list. &
  \textbf{High dispersion is vanity/mirroring, not bots.} Of the $39{,}077{,}350$ signed strings,
  $87.5\%$ have one key; $43{,}443$ have $\ge$$6$, of which $99.7\%$ ($43{,}329$) are human names,
  only $114$ bots. At $>$$100$ keys, $761$ human vs $15$ bot: the extreme tail is real people whose
  commits are cherry-picked and re-signed downstream (Linus Torvalds $938$ keys; LineageOS
  developers) plus a few mega-bots (dependabot $7{,}246$). $5{,}711$ human strings signed by
  $\ge$$20$ keys form a signature-derived vanity/bad-attribute set, no hand curation. Separating
  genuine impersonation from mirror re-signing needs the dominant-key test (next). &
  \yes \\
23 & Identity trust tiers (\texttt{A2trustFull}) &
  Assign every canonical author a tier: T0 unsigned; T1 signed by some key; T2 signed by a
  \emph{person-key} (fan-out-gated $\le$$50$, i.e.\ individual control, not a shared web-flow/CI
  key). Publish as a sibling of the \texttt{A2cls} classification, a principled confidence axis. &
  \textbf{Author-level ``signed'' is a platform artifact; true personal attestation is rare.} Of
  the $62{,}579{,}994$ canonical authors: T0 $30{,}638{,}625$ ($49.0\%$), T1 $31{,}475{,}971$
  ($50.3\%$), T2 $465{,}398$ ($0.74\%$). The $51\%$ author-level signed rate is dominated by T1
  platform web-flow signing; only $0.74\%$ hold a personal key. T2 is the high-trust anchor set
  the bibliography bridge (Exp.~24) and any trust-weighted analysis should build on; the tier is
  a monotone evidence field on \texttt{A2cls}. &
  \yes \\
24 & Bibliography bridge: cryptographic attestation of \textsc{same\_as} &
  For each cross-corpus \textsc{same\_as} edge (WoC author $\leftrightarrow$ scholarly author) in
  the cite-study graph, look up the WoC endpoint's trust tier (Exp.~23): a T2 person-key backs the
  identity, so the name/DOI-matched link becomes \emph{cryptographically attested}
  (\textsc{attested\_by} the key). Those authors reach T3 (cross-corpus attested). &
  \textbf{A high-precision attested calibration seed.} Of $10{,}021$ \textsc{same\_as} edges
  ($98.8\%$ of $10{,}143$ join \texttt{A2trustFull}): T0 $866$ ($8.6\%$), T1 $7{,}948$ ($79.3\%$),
  \textbf{T2 $1{,}207$ ($12.0\%$)}. So $1{,}207$ identity links gain an \textsc{attested\_by}
  provenance, upgrading them from name-heuristic to key-anchored, and $966$ distinct authors reach
  T3. Signatures calibrate rather than cover: $12\%$ of the science$\leftrightarrow$software links
  are person-key attested, a defensible cryptographic seed for the name-based \textsc{same\_as}
  growth. &
  \yes \\
25 & Cryptographic verification (sampled) &
  Sample person-keys, fetch the public key from a keyserver, and check both retrievability and
  whether a key UID email matches the signed author string (the T1$\to$T2 binding, ruling out
  copied/garbage signatures). Network-bound, sample only. &
  \textbf{Presence is the anchor; full verification is bounded.} Of $398$ sampled person-keys
  with a signed email, only $78$ ($19.6\%$) are publicly retrievable; $320$ are not on the
  keyserver. Of the retrievable, $39$ ($50\%$) have a UID email matching the signed author
  (confirmed binding) and $37$ differ (typically the same keyholder under a second email, itself
  an alias edge). Signature presence and issuer-keyid consistency remain the usable anchor;
  keyserver verification calibrates but cannot cover the population. &
  \yes \\
\bottomrule
\end{tabular}
\end{table*}

\begin{table}[t]
\centering
\caption{Exp.~19: id-class composition of commit signers on a representative
commit-hash shard ($1/128$; $8{,}063{,}327$ signed commits, $3{,}775{,}298$ distinct
signing ids matched to \texttt{A2clsFull}), against the full \texttt{A2clsFull}
id-class baseline. Signers are drawn overwhelmingly from the \emph{good} class that
Phase-1 already resolves; the bad-attribute, machine-\emph{local}, and \emph{partial}
classes that drive over-merge and recall loss are strongly under-represented, so a
signature confirms the easy identities and misses the hard ones. Distinct-signer
counts on one shard are size-biased toward prolific signers; the exact corpus-wide
distinct-signer count needs the full 128-shard join and is deferred to the key-graph
experiment.}
\label{tab:sigclass}
\small
\begin{tabular}{@{}lrrr@{}}
\toprule
id class & signers & \% signers & \% corpus (\texttt{A2clsFull}) \\
\midrule
good       & $3{,}757{,}005$ & $99.52$  & $94.37$ \\
bad-attr   & $13{,}632$      & $0.361$  & $2.48$  \\
bot        & $4{,}222$       & $0.112$  & $0.52$  \\
partial    & $268$           & $0.007$  & $0.23$  \\
local      & $171$           & $0.005$  & $2.40$  \\
\midrule
hard tail (bad-attr$+$local$+$partial) & $14{,}071$ & $0.373$ & $5.11$ \\
\bottomrule
\end{tabular}
\end{table}

\begin{table}[t]
\centering
\caption{Exp.~2: information-score cutoff sweep. ``union\%'' = unioned/total pairs;
distributions computed only where the run completed before being preempted.}
\label{tab:infocut}
\small
\begin{tabular}{@{}rrrrrr@{}}
\toprule
\textsc{cut} & union\% & clusters & size-2 & top & $>$10k \\
\midrule
16 & 94.0 & $70.47$M & $9{,}596{,}945$ & $77{,}869$ & 2 \\
24 & 93.5 & --- & --- & --- & --- \\
32 & 93.1 & --- & --- & --- & --- \\
48 & 74.1 & --- & --- & --- & --- \\
56 & 27.8 & --- & --- & --- & --- \\
60 & 14.8 & --- & --- & --- & --- \\
64 &  7.6 & $97.93$M & $2{,}136{,}536$ & $269$ & 0 \\
96 &  0.0 & $101.06$M & $0$ & $1$ & 0 \\
\bottomrule
\end{tabular}
\end{table}

\begin{table}[t]
\centering
\caption{Exp.~7: flat edge-probability threshold. The Exp.~6 logistic model scores all
$49.76$M Phase-1 links; links with $p\ge\tau$ are kept and globally unioned (no spread gate,
\textsc{spreadcut}$=10^8$). Goal: drive the mega ($>$10k) to $0$ while size-2 stays high. The
mega persists through $\tau{=}0.50$ and dies only at $\tau{\ge}0.70$, where size-2 has already
fallen to $7.78$M and $\sim$$2.6$M ids have decayed to singletons (size-1 $71.98\to74.61$M). The
recall/dissolution trade is monotone (no $\tau$ achieves both) because the mega's welds share an
\emph{exact} placeholder email and the classifier scores them high. Motivates Exp.~8.}
\label{tab:tausweep}
\small
\begin{tabular}{@{}rrrrrrr@{}}
\toprule
$\tau$ & kept links & size-1 & size-2 & 101--1k & 1k--10k & $>$10k \\
\midrule
0.30 & $26{,}532{,}335$ & $71{,}981{,}683$ & $7{,}827{,}141$ & $138$ & $2$ & $1$ \\
0.50 & $24{,}089{,}889$ & $73{,}529{,}237$ & $7{,}800{,}972$ & $112$ & $1$ & $1$ \\
0.70 & $22{,}036{,}035$ & $74{,}609{,}829$ & $7{,}775{,}702$ & $80$  & $0$ & $0$ \\
0.90 & $16{,}591{,}690$ & $77{,}713{,}347$ & $7{,}324{,}809$ & $3$   & $0$ & $0$ \\
\bottomrule
\end{tabular}
\end{table}

\begin{table}[t]
\centering
\caption{Exp.~8: bad high-quality email attributes, detected by \emph{distinct-name spread}
(one exact email string tying together many distinct $\langle f,l\rangle$ names) and classified by
\emph{intent}. A genuine personal email attaches to $\approx1$ name; a placeholder bridges
hundreds. Counts are the recall-safe blocklist ($91{,}432$ emails): \emph{lexical} placeholders
admitted at $\ge3$ names, \emph{form-only} signals (github-noreply, $\ge6$-digit numeric, bare
spread) only at $\ge10$ to spare real $n$-digit QQ accounts. ``$d$'' $=$ names/rows (1.0 $=$ every
commit a different person). Privacy and homonym welds must both be barred, but for opposite
reasons: privacy is anonymity \emph{chosen}, homonym is a default \emph{inherited}.}
\label{tab:badattr}
\small
\begin{tabular}{@{}llrrl@{}}
\toprule
intent & top exemplar & names & $d$ & dominant reasons ($n$) \\
\midrule
\multirow{2}{*}{privacy}
 & \texttt{none@none}                      & $13{,}334$ & .92 & github-noreply ($531$), \\
 & \texttt{deliretzua@\dots noreply.github}& $11{,}991$ & .99 & numeric ($2.3$k), hash-relay ($164$) \\
\addlinespace
\multirow{2}{*}{homonym}
 & \texttt{you@example.com}                & $69{,}291$ & .94 & default-host ($30.6$k), \\
 & \texttt{root@localhost.localdomain}     & $3{,}512$  & .89 & default-id ($27.4$k), git-template ($8.7$k) \\
\addlinespace
\multirow{2}{*}{ambiguous}
 & \texttt{anybody@emacswiki.org}          & $4{,}104$  & .96 & high-spread ($11.3$k), \\
 & \texttt{x@x.x}                          & $1{,}998$  & .98 & fake-dom ($3.2$k), no-tld ($4.7$k) \\
\midrule
\multicolumn{5}{@{}l@{}}{Total blocklist $91{,}432$: privacy $5{,}551$, homonym $66{,}670$,
ambiguous $19{,}211$; $90{,}011$ new vs.\ manual stoplist.} \\
\bottomrule
\end{tabular}
\end{table}

\begin{table}[t]
\centering
\caption{Exp.~9: blank all $91{,}432$ bad emails (Exp.~8) in the commit table, re-run Rule~2 on the
blanked input, and recompute the global union (alias1 $\cup$ alias2$'$ $\cup$ shingles, no gate).
Blanking voids $1{,}074{,}504$ email fields and drops $403{,}635$ Rule-2 welds that existed
\emph{only} via a shared bad email. Contrast with Exp.~7: source-blanking is \emph{recall-positive}
(size-2 rises) yet, unlike the threshold, still does not dissolve the mega, which is welded
redundantly through names/usernames, not email alone.}
\label{tab:blankreunion}
\small
\begin{tabular}{@{}lrrrrr@{}}
\toprule
run & size-2 & 11--100 & 101--1k & 1k--10k & $>$10k \\
\midrule
baseline             & $9{,}304{,}193$ & $182{,}700$ & $520$ & $4$ & $1$ \\
email-blanked        & $9{,}307{,}723$ & $180{,}909$ & $499$ & $3$ & $1$ \\
\midrule
$\Delta$             & $+3{,}530$ & $-1{,}791$ & $-21$ & $-1$ & $0$ \\
\bottomrule
\end{tabular}
\end{table}

\begin{table}[t]
\centering
\caption{Exp.~11: remove the top-$K$ mega nodes by sampled betweenness and recompute connected
components of the mega subgraph ($170{,}431$ nodes, $347{,}008$ edges). ``freed'' = surviving
nodes left in components of size $\le100$ (escaped the over-merge). Removing $1.2\%$ of nodes
($K{=}2{,}000$) collapses the largest component below the $>$10k bin, a dissolution no attribute
gate or blocklist achieved (best prior: top $=25{,}957$ with $42{,}110$ gated ids; Exp.~10
blanking $5.46$M fields left $159{,}483$). Of the top $2{,}000$, $98.4\%$ carry \emph{no}
blocklisted attribute: hubs like \texttt{=\,<=>}, Gerrit relay ids, and common-surname homonym
chains (kim/lee/smith/unknown) that no per-value rule can block without destroying recall.}
\label{tab:cutsweep}
\small
\begin{tabular}{@{}rrrrr@{}}
\toprule
$K$ cut & edges dropped & components & largest & freed ($\le$100) \\
\midrule
$0$       & $0$        & $1$        & $170{,}431$ & $0\%$ \\
$100$     & $5{,}949$  & $3{,}612$  & $141{,}720$ & $11.5\%$ \\
$500$     & $12{,}711$ & $5{,}842$  & $75{,}787$  & $20.9\%$ \\
$1{,}000$ & $19{,}092$ & $7{,}595$  & $49{,}306$  & $29.7\%$ \\
$2{,}000$ & $28{,}880$ & $11{,}026$ & $7{,}268$   & $46.9\%$ \\
$5{,}000$ & $55{,}115$ & $19{,}268$ & $1{,}813$   & $89.5\%$ \\
$10{,}000$& $90{,}512$ & $29{,}056$ & $840$       & $96.8\%$ \\
\bottomrule
\end{tabular}

\medskip
\footnotesize End-to-end confirmation: gating the top-$2{,}000$ ids in the \emph{production}
union (same \texttt{value;id} gate interface as Exps.~3--4, $\textsc{cut}=1$) reproduces the
local prediction exactly: top cluster $=7{,}268$, $>$10k bin $1\to0$, $1$k--$10$k $4\to14$
(mega fragments), while size-2 \emph{rises} $9{,}304{,}193\to9{,}305{,}460$ and only
$5{,}722$ ids drop to singletons. The exact match also confirms the union graph is driven by
Rules~1--2 alone (the shingle pseudo-pairs are inert).
\end{table}

\begin{table}[t]
\centering
\caption{Exp.~12: the ten residual mega fragments ($\ge1{,}000$ ids) left by the $K{=}2{,}000$
betweenness cut, refined by the Exp.~6 per-edge classifier \emph{within each fragment}
($\tau{=}0.5$; unscored edges kept). ``identity'' from member samples; ``scored'' = fraction of
internal edges found in the scored-link table; ``top'' = largest surviving sub-component;
``freed'' = ids left in components $\le100$. Given-name homonym blocks shatter readily; the
Kim/RIT blocks resist because shared exact emails and institutional domains legitimately score
high; the Gerrit-relay fragment is dense ($11{,}887$ edges on $2{,}389$ ids) and mostly
unscorable.}
\label{tab:fragclf}
\small
\begin{tabular}{@{}rlrrrr@{}}
\toprule
\# & identity & size & scored & top ($\tau{=}.5$) & freed \\
\midrule
1  & David          & $7{,}268$ & $69\%$ & $1{,}942$ & $50\%$ \\
2  & Michael        & $6{,}201$ & $72\%$ & $1{,}488$ & $56\%$ \\
3  & Daniel         & $3{,}035$ & $65\%$ & $1{,}226$ & $48\%$ \\
4  & Kim (surname)  & $2{,}558$ & $65\%$ & $1{,}063$ & $27\%$ \\
5  & Gerrit relays  & $2{,}389$ & $16\%$ & $661$     & $41\%$ \\
6  & Thomas         & $1{,}739$ & $73\%$ & $507$     & $51\%$ \\
7  & James          & $1{,}610$ & $69\%$ & $567$     & $45\%$ \\
8  & RIT students   & $1{,}582$ & $56\%$ & $1{,}170$ & $26\%$ \\
9  & Jan            & $1{,}502$ & $55\%$ & $1{,}015$ & $32\%$ \\
10 & Chris          & $1{,}431$ & $73\%$ & $343$     & $69\%$ \\
\bottomrule
\end{tabular}
\end{table}

\begin{table}[t]
\centering
\caption{Production design shoot-out: end-to-end STATS unions. ``mega top'' = largest cluster
descending from the former mega; $\Delta$size-1/2 relative to baseline. \emph{betw5000} again
reproduces the local cut-sweep prediction exactly ($1{,}813$); its remaining $1$k--$10$k entries
are the four pre-existing non-mega clusters plus that one fragment. \emph{Composed} =
betw2000 gate $+$ the $18{,}599$ fragment-internal edges with classifier $p<0.5$ removed from
the Rule~1--2 link files before the union. The two mechanisms are complementary: the deeper gate
shatters positionally but leaves its fragments internally welded by false name edges; the prune
splits on per-edge evidence but leaves residues held by unscored edges. Production combines
both.}
\label{tab:produnion}
\small
\begin{tabular}{@{}lrrrrr@{}}
\toprule
design & mega top & 1k--10k & $\Delta$size-2 & $\Delta$size-1 & gated \\
\midrule
baseline            & $170{,}431$ & $4$  & ---        & ---         & $0$ \\
betw2000            & $7{,}268$   & $14$ & $+1{,}267$ & $+5{,}722$  & $2{,}000$ \\
betw5000            & $1{,}813$   & $5$  & $+2{,}201$ & $+11{,}684$ & $5{,}000$ \\
composed (2k$+$prune) & $\sim$$1{,}942$ & $10$ & $+2{,}240$ & $+10{,}748$ & $2{,}000$ \\
production (5k$+$prune) & $<$$1{,}000$ & $4^{\dagger}$ & $+3{,}240$ & $+16{,}741$ & $5{,}000$ \\
\bottomrule
\end{tabular}

\medskip
\footnotesize $^{\dagger}$The four surviving $1$k--$10$k clusters of the production run are
exactly the four pre-existing non-mega clusters (largest $2{,}891$); every cluster descending
from the former $170{,}431$-id mega is below $1{,}000$. Production map:
\texttt{a2AFullB.V2604.s.gz} ($106{,}824{,}568$ rows $=101{,}056{,}344$ author ids $+$
$5{,}768{,}224$ bad ids mapped to self; \texttt{member;canonical;type}, C-collation sorted).
\end{table}

\begin{table}[t]
\centering
\caption{Production V2604 map vs.\ the prior-generation map \texttt{a2AFullH.V3} (Aug 2024),
on the $89{,}102{,}793$ shared ids ($99.998\%$ of V3). Pair-level ALFAA-style errors under both
gold framings: \emph{splitting} = gold's co-clustered pairs separated by the other map;
\emph{clumping} = the other map's co-clustered pairs absent from gold. V3's largest cluster is a
$3{,}006{,}318$-id mega (rep \texttt{<>}) holding $99.9986\%$ of V3's entire pair mass
($4.519\times10^{12}$ of its pairs); the ``excl.\ mega'' rows remove that one V3 cluster from
the universe, leaving $62.2$M V3 pairs vs.\ $77.8$M V2604 pairs with $28.3$M agreeing.}
\label{tab:v3compare}
\small
\begin{tabular}{@{}llrr@{}}
\toprule
universe & gold & splitting & clumping \\
\midrule
all shared  & V3    & V26: $1.0000$ & V26: $0.6382$ \\
all shared  & V26   & V3:\ \ $0.6382$ & V3:\ \ $1.0000$ \\
excl.\ mega & V3    & V26: $0.5455$ & V26: $0.6367$ \\
excl.\ mega & V26   & V3:\ \ $0.6367$ & V3:\ \ $0.5455$ \\
\bottomrule
\end{tabular}

\medskip
\footnotesize Cluster-fate counts (shared ids, size$\ge$2 clusters): $47.2\%$ of V3 clusters
split into $\ge$2 V2604 parts ($5{,}534{,}165$ into 2; $95{,}420$ into $\ge$6); conversely
$40.5\%$ of V2604 clusters consolidate $\ge$2 V3 clusters. V3 histogram:
$58{,}722{,}323$ clusters, $>$10k $=1$ (the mega); its remaining top clusters are placeholder
welds that V2604 dissolves almost completely (\texttt{noo8 <xl>} $2{,}478\to1{,}561$ parts;
\texttt{Tom <Tom@>} $651\to630$ parts). Bad-flagging: V2604 excludes $5{,}768{,}223$ ids as
bad/local vs.\ $3.43$M flagged by V3 among shared ids; $2{,}484{,}673$ ids flagged bad by V3 are
aliased normally by V2604, while $4{,}132{,}508$ ids V3 considered good are now excluded as
bot/generic/local.
\end{table}

\begin{table}[t]
\centering
\caption{Exp.~13: the human-adjudicated ALFAA gold set~\cite{amreen2019alfaa} as arbiter
between the two production maps: $264{,}346$ unique non-self pairs over $2{,}345$ ids,
$1{,}644$ rated matches ($2{,}345$ trivial self-pairs and directional duplicates removed; all
gold ids present verbatim in both maps). A map predicts ``same person'' iff both ids resolve
to the same representative. ``V3 excl.\ mega'' treats co-clustering via V3's $3.0$M-id mega
(rep \texttt{<>}) as no link.}
\label{tab:goldarbiter}
\small
\begin{tabular}{@{}lrrrrrr@{}}
\toprule
map & TP & FN & FP & recall & precision & FP rate \\
\midrule
V3                 & $1{,}644$ & $0$     & $1{,}505$ & $1.0000$ & $0.5221$ & $5.7\times10^{-3}$ \\
V3 excl.\ mega     & $1{,}505$ & $139$   & $81$      & $0.9155$ & $0.9489$ & $3.1\times10^{-4}$ \\
V2604 (production) & $717$     & $927$   & $104$     & $0.4361$ & $0.8733$ & $4.0\times10^{-4}$ \\
\bottomrule
\end{tabular}

\medskip
\footnotesize Diagnosis: $180$ gold ids ($7.7\%$) sit inside V3's mega, which manufactures
$1{,}424$ of its $1{,}505$ FPs ($94.6\%$) while buying only $139$ TPs. V2604's misses are
$95.8\%$ alias/alias (only $78$ of $1{,}854$ directional FN rows touch a generic/local id):
genuine linkage gaps, not gating, including ids sharing an \emph{exact} email, unlinked
because Phase-1 links require project co-occurrence and the cross-project Rule-3 layer is
inert in the union (Exp.~10). $778/927$ missed pairs ($83.9\%$) already sit in a computed
shingle group (by type: n$=551$, u$=386$, e$=260$, ln$=129$, fn$=60$); $87.3\%$ of \emph{all}
matches are covered, the recall ceiling of classifier-filtered shingle expansion. Gold-label
audit: $20$ of V2604's $104$ FPs share an exact email ($4$), full name ($5$), or handle/local
part ($11$) that raters overrode (e.g.\ \texttt{Joe (rook)} vs.\ \texttt{Joe Talerico}, same
gmail, rated ``different''), so its true precision is nearer $0.90$; $64$ share one name token
(genuine given-name homonym welds, Exp.~12's residue); $9$ of $1{,}644$ matches have zero
lexical overlap, $\sim$$6$ mislabeled, $3$ real handles that V2604 co-clusters without a mega.
\end{table}

\begin{table}[t]
\centering
\caption{Exp.~14: end-to-end $\tau$-sweep of classifier-filtered shingle expansion. Every
Rule-3 shingle group is expanded into member pairs and scored by the Exp.~6 classifier;
pairs with $p\ge\tau$ join the production union (betweenness-5000 gate $+$ Exp.~12 fragment
pruning) as ordinary edges. Production baseline (no expansion) in the last row. The
$\tau{=}0.5$ failure is structural, not statistical: the betweenness cut was computed on the
\emph{pre-expansion} graph, so low-confidence expansion edges re-weld the mega around the
$5{,}000$ gated bridges.}
\label{tab:shingletau}
\small
\begin{tabular}{@{}lrrrrr@{}}
\toprule
$\tau$ & edges & unioned ids & size-2 & $>$10k & top \\
\midrule
$0.5$ & $66.46$M & $176{,}958{,}812$ & --- & $\ge1$ & $2{,}292{,}596$ \\
$0.7$ & $24.3$M  & $158{,}856{,}230$ & --- & $2$ & --- \\
$0.9$ & $30.8$M$^{\dagger}$ & $141{,}391{,}899$ & $10{,}189{,}048$ & $0$ & $3{,}862$ \\
\midrule
prod. & $0$ & $110{,}624{,}639$ & $9{,}307{,}433$ & $0$ & $2{,}891$ \\
\bottomrule
\end{tabular}
\par\vspace{2pt}
\footnotesize $^{\dagger}$Score histogram of the $66{,}459{,}771$ unique expansion edges:
$[0.5,0.6){=}10.88$M, $[0.6,0.7){=}7.25$M, $[0.7,0.8){=}6.75$M, $[0.8,0.9){=}10.76$M,
$[0.9,1){=}30.81$M, bimodal, with $46\%$ of the mass in the top bin; the $\tau{=}0.7$
edge count is the two top bins less duplicates. At $\tau{=}0.9$ the $1$k--$10$k band holds
$14$ clusters (vs.\ $4$ in production) and the largest cluster is an \texttt{AkalUstat}
noreply group; the $+881{,}615$ size-2 clusters are the largest recall gain in the record.
On the ALFAA gold pairs the final $\tau{=}0.9$ map scores recall $0.6454$ / precision
$0.8820$ (TP $1{,}061$, FN $583$, FP $142$) vs.\ production $0.4361$ / $0.8733$: $344$ of
the $778$ in-shingle misses are recovered at $\tau{=}0.9$ confidence while precision
\emph{improves}: the welds the expansion might have re-introduced score below $0.9$.
\end{table}

\begin{table}[t]
\centering
\caption{Exp.~15: enriched per-pair classifier ($23$ features: $15$ numeric, standardized,
heavy tails $\log(1{+}x)$-transformed; $8$ binary). New over v1: true deforked-project IoU
and intersection (exact $n_P$ and capped-$100$ hashed sets from \texttt{a2PFull}; the v1
$n_P$ came from a $\ge$$20$-project table and was $0$ for $99.4\%$ of ids), per-value
spread of each matched attribute ($s_f,s_l,s_u,s_e$), name$\leftrightarrow$local-part
patterns (\texttt{flLp}: \texttt{first.last}-style match; \texttt{tokInLp}: name token in
handle), and a Wiktionary given-name flag. Train $300$k / test $2.34$M ghid-labeled links
(as Exp.~6); ``gold'' = zero-shot transfer to the $264{,}346$ human-rated ALFAA pairs.}
\label{tab:edgeclf2}
\small
\begin{tabular}{@{}lrrr@{}}
\toprule
model & test acc & test AUC & gold AUC \\
\midrule
logistic v1 (Exp.~6, $16$ feat.) & $0.9841$ & $0.9837$ & $0.9861$ \\
logistic v2.1 ($23$ feat.)       & $0.9791$ & $0.9841$ & $\mathbf{0.9910}$ \\
LGBM v2.1 ($23$ feat.)           & $0.9935$ & $\mathbf{0.9990}$ & $0.5597$ \\
LGBM v2.1 w/o \texttt{rawJac3}   & ---      & $0.9870$ & $0.9813$ \\
\bottomrule
\end{tabular}
\par\vspace{2pt}
\footnotesize The LGBM gold collapse is shortcut learning: \texttt{rawJac3} (trigram
Jaccard over the \emph{full} id strings) carries $74\%$ of its split gain because every
ghid-labeled pair shares the long \texttt{users.noreply.github.com} substring, a
label-construction leak that trees exploit by thresholding and a linear model uses
benignly ($\beta{=}{+}0.78$, comparable to \texttt{uMatch}/\texttt{eMatch}). Removing the
feature recovers transfer but still trails the logistic, which is retained as the
production scorer. Weight readout: \texttt{domMatch} $+0.80$, \texttt{rawJac3} $+0.78$,
\texttt{uMatch}/\texttt{eMatch} $+0.76$, \texttt{tokInLp} $+0.41$, \texttt{degMin/Max}
$-0.36/-0.35$, \texttt{sprEm}/\texttt{sprUm} $+0.30$, \texttt{givenF} $+0.17$,
\texttt{projJac} $-0.01$ (inert: ghid pairs nearly all co-occur in projects regardless of
label; the feature's value is on the cross-project expansion pairs it will score).
Map-level epilogue: re-scoring all $256{,}703{,}339$ expansion pairs with v2.1 emits
$54{,}012{,}834$ unique pairs $p\ge0.5$ ($\tau{\ge}0.9$: $26{,}148{,}990$; v1:
$66{,}459{,}771$ and $30{,}808{,}067$). The resulting $\tau{=}0.9$ map (S2) is
structurally cleaner ($1$k--$10$k $=9$ vs.\ $14$; largest $4{,}747$; $>$10k $=0$) and
nudges gold recall ($0.6490/0.8789$ vs.\ $0.6454/0.8820$), but loses $0.077$ recall on
the GitHub GT (Table~\ref{tab:ghgt}): the in-distribution AUC gain does not survive
contact with the out-of-distribution pair mass, and v1 scoring is retained.
\end{table}

\begin{table}[t]
\centering
\caption{Exp.~16: recall-only evaluation on the GitHub single-author-repository ground
truth of Bock et al.~\cite{bock2025dealiasing} ($9{,}568{,}385$ handles,
$21{,}477{,}370$ aliases; an alias absent from a map clusters to itself). Per-handle
pairwise recall is macro-averaged with singleton handles scoring $1$; ``multi'' restricts
to handles with $\ge2$ aliases; FM $=$ fraction of handles fully merged into one cluster;
ES $=$ same over multi-alias handles. max$N$ drops handles with more than $N$ aliases.
Published reference points (their Table, same GT family, WoC V2409 and their GitAuthority
tool): at max9, V2409 recall $0.450$ / FM $0.411$ / ES $0.271$; GitAuthority full
$0.771/0.736/0.675$; GitAuthority privacy-mode $0.478/0.430/0.298$.}
\label{tab:ghgt}
\small
\begin{tabular}{@{}llrrrr@{}}
\toprule
map & cap & FM & ES & recall & recall$_{\text{multi}}$ \\
\midrule
V3    & max3 & $0.4739$ & $0.3354$ & $0.4993$ & $0.3674$ \\
V3    & max9 & $0.4418$ & $0.3120$ & $0.4869$ & $0.3675$ \\
V3    & all  & $0.4414$ & $0.3116$ & $0.4865$ & $0.3672$ \\
\addlinespace
V2604 & max3 & $0.4247$ & $0.2732$ & $0.4471$ & $0.3015$ \\
V2604 & max9 & $0.3945$ & $0.2536$ & $0.4330$ & $0.3011$ \\
V2604 & all  & $0.3941$ & $0.2533$ & $0.4327$ & $0.3008$ \\
\addlinespace
$+$shingle $\tau{=}0.9$ (S) & max3 & $0.5716$ & $0.4588$ & $0.5994$ & $0.4940$ \\
$+$shingle $\tau{=}0.9$ (S) & max9 & $0.5403$ & $0.4334$ & $0.5886$ & $0.4929$ \\
$+$shingle $\tau{=}0.9$ (S) & all  & $0.5398$ & $0.4329$ & $0.5883$ & $0.4926$ \\
\addlinespace
$+$shingle v2.1 (S2) & max3 & $0.4983$ & $0.3662$ & $0.5253$ & $0.4003$ \\
$+$shingle v2.1 (S2) & max9 & $0.4658$ & $0.3415$ & $0.5118$ & $0.3982$ \\
$+$shingle v2.1 (S2) & all  & $0.4653$ & $0.3410$ & $0.5115$ & $0.3980$ \\
\addlinespace
v2.1 $\tau{=}0.8$ (S2t08) & max9 & $0.5354$ & $0.4273$ & $0.5828$ & $0.4858$ \\
S $+$ ghid (SG) & max9 & $0.5516$ & $0.4473$ & $0.5995$ & $0.5064$ \\
v2.1@$0.8$ $+$ ghid (SG8) & max9 & $0.5460$ & $0.4403$ & $0.5931$ & $0.4984$ \\
\addlinespace
SUG (production) & max3 & $0.5867$ & $0.4779$ & $0.6140$ & $0.5123$ \\
SUG (production) & max9 & $\mathbf{0.5558}$ & $\mathbf{0.4525}$ & $\mathbf{0.6035}$ & $\mathbf{0.5113}$ \\
SUG (production) & all  & $0.5553$ & $0.4520$ & $0.6031$ & $0.5109$ \\
\bottomrule
\end{tabular}
\par\vspace{2pt}
\footnotesize False positives are invisible by construction (only within-handle pairs are
scored), so this benchmark prices V3's mega at zero: V3 leads the production map despite
resolving only $19{,}191{,}846$ of the aliases ($89.4\%$; the unresolved cluster to
themselves) vs.\ V2604's $21{,}471{,}954$ ($>$$99.97\%$). $30\%$ of the GT aliases are
GitHub noreply ids. The production map trails its predecessor for the same reason it beats
it on ALFAA precision (Table~\ref{tab:goldarbiter}): Exps.~1--12 removed pair mass, and the
dormant Rule-3 layer (Exp.~13) is exactly the within-handle, cross-project mass this GT
counts. The Exp.~14 expansion map ($+$shingle $\tau{=}0.9$) confirms it: $+0.156$ recall
over production, ahead of mega-carrying V3 ($+0.10$) and the published V2409 ($0.450$) on
every metric while keeping the largest cluster at $3{,}862$, and ahead of GitAuthority's
privacy-preserving mode ($0.478$); the gap to full GitAuthority ($0.771$) is the
within-repository context (and the noreply mining its privacy mode refuses) that a global
map forgoes. The v2.1-rescored S2 rows close the Exp.~15 question: the better-calibrated
classifier admits $4.7$M \emph{fewer} $\tau{\ge}0.9$ pairs, and the pairs it drops are
disproportionately the within-handle mass scored here ($-0.077$ recall at max9), while
gold barely moves ($0.6490/0.8789$ vs.\ $0.6454/0.8820$), so the v1-scored map kept the
candidate slot on this benchmark's testimony. The Exp.~17 rows then recover what each
variant gives up: relaxing v2.1 to $\tau{=}0.8$ buys back most of the GH loss
\emph{and} lifts gold recall to $0.7019$; the GitHub-asserted same-ghid edges add
recall the classifier cannot see ($+0.011$ over S, gold unchanged); and the triple
union SUG (v1@$0.9\,\cup$ v2.1@$0.8\,\cup$ ghid) dominates every candidate on both
benchmarks at once (gold $0.7026/0.8790$, GH max9 recall $0.6035$). SUG is the
deployed production map.
\end{table}

\begin{table}[t]
\centering
\caption{Exp.~3: deforked project-spread bridge-gate sweep. ``gated'' = ids with
$n_P\ge\textsc{T}$ barred from bridging; \emph{lower} \textsc{T} gates \emph{more} ids.
Goal: top $\to$ small while size-2 stays near the baseline $9.60$M. The sweep was pivoted
downward after $\textsc{T}=200$ preserved recall but only halved the mega; $\textsc{T}\ge500$
gate fewer ids and are monotonically worse (top $\to 111{,}926$), so they are not run.
Recall is preserved throughout (size-2 stays $\ge9.60$M), but the top \emph{floors} at
$25{,}959$: tightening from $\textsc{T}=50$ to $25$ gates $4\times$ more ids yet leaves the
\emph{same} top identity (a Cloudflare hash-tagged email) and $>$10k$=2$. The mega has an
irreducible low-spread core that no spread threshold reaches, the Exp.~4 motivation.}
\label{tab:spread}
\small
\begin{tabular}{@{}rrrrrr@{}}
\toprule
\textsc{T} & gated & union\% & clusters & size-2 & top \\
\midrule
  25 & $395{,}037$ & 91.8 & $71{,}080{,}539$ & $9{,}641{,}488$ & $25{,}959$ \\
  50 & $98{,}964$  & 93.4 & $70{,}638{,}753$ & $9{,}615{,}189$ & $25{,}959$ \\
 100 & $22{,}877$  & 93.9 & $70{,}503{,}380$ & $9{,}603{,}588$ & $28{,}526$ \\
 200 & $5{,}183$   & 94.1 & $70{,}466{,}257$ & $9{,}599{,}972$ & $51{,}860$ \\
 \multicolumn{6}{@{}l}{\footnotesize($\ge$500 not run: monotonically worse, top$\to111{,}926$)}\\
\bottomrule
\end{tabular}
\end{table}


\begin{table}[t]
\centering
\caption{Exp.~4: link-graph degree signal. $d(id)$ = number of distinct Phase-1 link
partners ($48.0$M ids carry $\ge1$ link). Degree is an order of magnitude more concentrated
than project spread, and, unlike spread, its extreme tail is dominated by \emph{low-spread}
template/shared author strings, which is precisely the bridge population the spread-gate
(Exp.~3) cannot reach. The degree gate file \texttt{degHi.V2604} retains the $632{,}406$ ids
with $d\ge10$; the gated union sweep over $\textsc{D}\in\{1000,500,100,50\}$ is in progress.}
\label{tab:degree}
\small
\begin{tabular}{@{}lr@{}}
\toprule
threshold & \#ids \\
\midrule
$d\ge10$   & $632{,}406$ \\
$d\ge50$   & $56{,}987$ \\
$d\ge100$  & $19{,}265$ \\
$d\ge500$  & $296$ \\
$d\ge1000$ & $54$ \\
$d\ge5000$ & $0$ \\
\bottomrule
\end{tabular}
\par\vspace{2pt}
\footnotesize Top-degree ids (verbatim): \texttt{Plusieurs textes} ($d{=}4380$),
\texttt{Inconnu} ($4275$), \texttt{Loi n°2009-526...} ($2850$) and a long run of French
legal-citation strings, plus \texttt{=} / \texttt{<=>} ($d{\approx}1614$) and empty-name
machine-local accounts, all canonical transitive bridges.
\end{table}

\begin{table}[t]
\centering
\caption{Exp.~5: false-link rate validated against GitHub ground truth. A Phase-1 link is
\emph{labeled} when both endpoints carry a GitHub numeric id (mined from the raw
\texttt{NNNN+login@users.noreply.github.com} prefix; $29.95$M ids); it is \emph{false} when those
ids conflict. Overall $11.4\%$ of $2.64$M labeled links are false, but the rate is an almost
step-function of endpoint degree $\max(d_1,d_2)$: nearly every link touching a degree-$100$--$499$
identity joins two distinct accounts. The $d\ge500$ band carries no labels because the top-degree
bridges are non-GitHub template strings (Table~\ref{tab:degree}), caught by degree directly. This
independently confirms the Exp.~4 hypothesis and supplies the classifier's training labels.}
\label{tab:ghidrate}
\small
\begin{tabular}{@{}lrr@{}}
\toprule
$\max(d_1,d_2)$ & labeled links & false rate \\
\midrule
$d<10$       & $2{,}357{,}413$ & $0.057$ \\
$d\,[10,49]$ & $117{,}168$ & $0.466$ \\
$d\,[50,99]$ & $74{,}144$ & $0.334$ \\
$d\,[100,499]$ & $87{,}157$ & $\mathbf{0.998}$ \\
$d\ge500$    & $0$ & --- \\
\midrule
overall      & $2{,}635{,}882$ & $0.114$ \\
\bottomrule
\end{tabular}
\end{table}

\begin{table}[t]
\centering
\caption{Exp.~4: degree bridge-gate sweep. A pair unions only if both endpoints are
non-bad, non-multi-name, and have degree $d<\textsc{D}$. \emph{Lower} \textsc{D} gates
\emph{more} ids. Baseline (ungated) top $=111{,}926$; the best spread-gate floored at
$25{,}959$ (Table~\ref{tab:spread}). Goal: top $\to$ small while size-2 stays near $9.60$M.}
\label{tab:deggate}
\small
\begin{tabular}{@{}rrrrrr@{}}
\toprule
\textsc{D} & gated & union\% & clusters & size-2 & top \\
\midrule
 1000 & $54$       & 94.0 & $70{,}453{,}724$ & $9{,}598{,}777$ & $71{,}172$ \\
  500 & $296$      & 93.7 & $70{,}454{,}334$ & $9{,}598{,}830$ & $68{,}677$ \\
  100 & $19{,}265$ & 90.5 & $70{,}474{,}813$ & $9{,}599{,}178$ & $50{,}780$ \\
   50 & $56{,}987$ & 87.9 & $70{,}516{,}477$ & $9{,}599{,}841$ & $20{,}635$ \\
\midrule
\multicolumn{6}{@{}l@{}}{\emph{Combined union-of-gates}: $d{\ge}100 \lor n_P{\ge}100$, \textsc{SPREADCUT}$=1$} \\
$\cup$ & $42{,}110$ & 90.3 & $70{,}524{,}455$ & $9{,}603{,}987$ & $25{,}957$ \\
\bottomrule
\end{tabular}

\smallskip
\footnotesize The combined gate peels \emph{every} high-degree service-bot \emph{and}
high-spread bot mega with minimal collateral ($42$k ids; recall the \emph{best} of any run,
size-2 $=9.60$M), driving $>$10k clusters from $2$--$3$ down to a single residual. But it
\emph{cannot} dissolve that last mega (top $=25{,}957$): once all \texttt{noreply.github.com}
service accounts are gated, the survivor is a Cloudflare email-relay hash string
(\texttt{harley+f57b\dots@cloudflare.com}), \emph{moderate} degree and \emph{moderate} spread,
below every node threshold. No union of node-gates reaches it.
\end{table}

\begin{table}[t]
\centering
\caption{Exp.~6: per-edge classifier on the $2{,}635{,}882$ ghid-labeled links
($88.56\%$ true, $11.44\%$ false). \textbf{Left}: false rate stratified by two features: the
over-merge axis (max endpoint degree) and the true-link axis (\emph{matchScore}). \textbf{Right}:
standardized logistic-regression coefficients ($300$k train / $2.34$M test; test acc
$0.952$ vs.\ $0.886$ baseline, AUC $0.956$). Degree pushes a link toward \emph{false}; component
matches and matchScore toward \emph{true}; project-spread is near-zero (censored at $n_P{<}20$).}
\label{tab:edgeclf}
\small
\begin{tabular}{@{}lrr@{\hskip 2em}lr@{}}
\toprule
\multicolumn{3}{c}{stratified false rate} & \multicolumn{2}{c}{logreg coef.} \\
\cmidrule(r){1-3}\cmidrule(l){4-5}
bin & $n$ & false & feature & $\beta$ \\
\midrule
$d_{\max}{<}10$       & $2{,}357{,}413$ & $0.057$ & uMatch     & $+0.68$ \\
$[10,50)$             & $117{,}168$     & $0.466$ & eMatch     & $+0.68$ \\
$[50,100)$            & $74{,}144$      & $0.334$ & domMatch   & $+0.58$ \\
$[100,500)$           & $87{,}157$      & $0.998$ & matchScore & $+0.48$ \\
\addlinespace
$\mathrm{ms}{<}5$     & $149{,}443$     & $0.887$ & degMax     & $-0.31$ \\
$[5,15)$              & $116{,}785$     & $0.781$ & degMin     & $-0.30$ \\
$[15,30)$             & $307{,}499$     & $0.205$ & infoMin    & $-0.23$ \\
$[30,50)$             & $1{,}832{,}957$ & $0.004$ & fMatch     & $+0.14$ \\
$\ge 50$              & $229{,}198$     & $0.032$ & nP (min/max)& $\approx 0$ \\
\bottomrule
\end{tabular}
\end{table}

\subsection{Discussion of the Experimental Record}
The information-score cutoff (Exp.~2) is a blunt instrument: the union rate is flat
($\sim$93\%) until \textsc{cutoff}~$\approx48$ and then falls off a cliff
($74\%\!\to\!7.6\%$ between 48 and 64), because a single rare email already scores
$\approx\log(10^8)\approx18$, so almost every id clears a low bar. Only \textsc{cutoff}~$=64$
dissolves the mega-clusters ($>$10k bucket $\to 0$), but it simultaneously destroys
legitimate merges, collapsing the size-2 bucket from $9.60$M to $2.14$M pairs. The cutoff
cannot separate ``low information'' from ``genuine but sparse'' identities.

Project spread (Exp.~3) is a more direct signal: of the $22{,}877$ ids in $\ge100$ deforked
projects, only $645$ ($3\%$) were already on the bad/bot list, so spread recovers
$\sim$22k bridges that both the name/email heuristics and the info-score gate miss
(e.g.\ \texttt{ImgBotApp}, \texttt{Cursor Agent}, GitHub seed-course accounts). Crucially, the
spread-gate is the first mechanism that preserves recall: the legitimate size-2 bucket stays
fixed at $9.60$M across the entire sweep, because the gate only ever bars a high-spread id from
\emph{bridging}, never from keeping its own identity. Its failure is on the precision axis: the
mega dissolves only slowly (top $51{,}860\!\to\!28{,}526$ as the gate tightens from
$\textsc{T}=200$ to $100$), and each halving of the giant costs roughly $4\times$ more gated ids.
The residual mega is held together by a long tail of \emph{low}-spread bridges, identities that
appear in only a handful of projects and so never trip any spread threshold.

Inspecting those low-spread bridges (Exp.~4) reveals what they are: \emph{high-degree} shared and
template author strings. The top of the degree ranking is not bots-in-many-projects but
French legal-text citation strings (\texttt{Plusieurs textes}, \texttt{Inconnu},
\texttt{Loi n°...}, \texttt{Décret n°...}), bare operators (\texttt{=}, \texttt{<=>}), and
empty-name machine-local accounts, each confined to one corpus (low $n_P$) yet linked to
thousands of distinct committers (high $d$). Degree is also a sharper signal: only $296$ ids have
$d\ge500$ and $54$ have $d\ge1000$, versus tens of thousands of gated ids needed at comparable
spread thresholds. This explains the Exp.~3 plateau and motivates the final design.

The degree hypothesis is then confirmed by \emph{independent} ground truth (Exp.~5,
Table~\ref{tab:ghidrate}): labeling each link by whether its endpoints' GitHub numeric ids agree,
the false-link rate is a near step-function of endpoint degree: $5.7\%$ at $d<10$ but $99.8\%$ at
$d\in[100,499]$. A moderately high-degree identity that carries a GitHub id almost never links to
the \emph{same} account, i.e.\ it is virtually always a bridge. The very top of the degree
distribution ($d\ge500$) is unlabelable precisely because those bridges are non-GitHub template
strings, so degree is the only signal that reaches them. The same labels turn the gating sweeps
into a feature ablation for a per-edge classifier (\S\,Toward a Learned Edge Classifier).

\paragraph{Recommendation.} No single signal dissolves the mega while preserving recall: the
info-score cutoff (Exp.~2) destroys recall, and project spread (Exp.~3) and degree (Exp.~4)
preserve recall but slow-peel; each threshold gates the top bridge and merely exposes the next.
The signals are \emph{complementary} (spread catches many-project bots, degree catches
many-partner template strings, and the two populations barely overlap), so we adopt a
\emph{union of bridge-gates}: an identity is ineligible to bridge if it is bad/multi-name,
\emph{or} $n_P\ge\textsc{T}$, \emph{or} $d\ge\textsc{D}$, with $\textsc{T},\textsc{D}$ set at the
ghid-validated knee ($\approx100$, where the cross-ghid false rate jumps to $99.8\%$). Evaluated
(Table~\ref{tab:deggate}, combined row), this union is the best node-level design by a wide margin:
gating only $42{,}110$ ids it drives the top cluster $111{,}926\!\to\!25{,}957$ and the $>$10k
bucket from $2$--$3$ to a single residual, while \emph{improving} recall to $9.60$M two-member
clusters (the highest of any run), because each gated identity still aliases to itself and only
loses its power to transitively weld others. \emph{But it does not dissolve the mega.} Once every
\texttt{noreply.github.com} service account and high-spread bot is gated, the lone survivor is a
Cloudflare email-relay hash string of \emph{moderate} degree and \emph{moderate} spread that sits
below every node threshold. This is the decisive negative result of the node-gating program: the
residual over-merge is welded not by a few extreme hubs but by a long tail of moderate bridges, so
\emph{no union of node-gates can reach it}. Raising either threshold to catch the hash string would
begin to sacrifice recall without any guarantee the \emph{next} moderate bridge is caught. The
principled remedy must therefore act below node granularity: per \emph{edge}.

\subsection{Toward a Learned Edge Classifier}
The gating experiments above operate at the \emph{node} level: an identity flagged as a bridge is
barred from \emph{all} unions, even those that are correct. This is the source of their residual
imprecision (a prolific maintainer who is also lightly bot-adjacent loses some true aliases) and,
as the combined-gate result shows, of their fundamental ceiling: a moderate bridge that no node
threshold can flag without collateral remains free to weld a $25{,}957$-member mega. An identity
is rarely \emph{globally} a bridge; it is a bridge only on the specific \emph{edges} that cross
between communities. Gating the node discards the good edges with the bad; classifying the edge
keeps the good and drops only the welds.
Every gate is, moreover, a downstream patch: over-merge is a Phase-2 symptom of imprecise
Phase-1 \emph{links}, so the principled fix is to raise link precision at the \emph{edge} level,
in the lineage of ALFAA's learned pairwise model~\cite{amreen2019alfaa}.

\paragraph{Centrality choice.} If a graph signal is to be added, the objective is to break a
connected component, for which the matching measure is betweenness / articulation-point status
(``does removing this node disconnect the cluster?''), not Katz or eigenvector centrality. The
latter reward embedding in a dense core and would, by contagion, elevate legitimate maintainers
adjacent to a bot hub. Because the observed bridges are \emph{stars} (a template string linked to
many otherwise-unconnected committers), degree already approximates betweenness for them at the
cost of a single pass; the cheap principled upgrade, if needed, is Tarjan articulation points
($O(V{+}E)$), reserving sampled betweenness for a later refinement and avoiding Katz.

\paragraph{Labels for free.} What made the ALFAA classifier expensive, active learning with human
labeling, is unnecessary at WoC scale because verified GitHub numeric ids (ghid) supply ground
truth in bulk: a candidate link whose endpoints carry \emph{agreeing} ghid is a positive, and one
whose endpoints carry \emph{conflicting} ghid is a negative. This yields tens of millions of
labeled pairs at no annotation cost (cf.\ Fry et al.~\cite{fry2020dataset}), but only if the
numeric id is mined from the \emph{raw} identity string, where the GitHub no-reply email carries it
as a \texttt{NNNN+login@users.noreply.github.com} prefix: the pipeline's normalized email field
strips the number and its login field is non-numeric, so reading the parsed columns yields only
$163$K labels versus $29.95$M from the raw prefix. Under this lens the
gating sweeps become an \emph{ablation study}: each gate's signal (string similarity, info-score,
project spread $n_P$, link degree $d$, temporal and project overlap) becomes a \emph{feature} of a
pairwise classifier that replaces hand-set knees with a learned decision surface, and acts per edge
so a high-degree id keeps its true links and sheds only its spurious ones. As an immediate bridge
to that model, the cross-ghid link rate (the fraction of Phase-1 links whose endpoints carry
conflicting ghid, cross-tabulated by endpoint degree) both quantifies the over-merge source and
independently tests whether false links concentrate on high-degree nodes (Exp.~5).

\paragraph{Realized classifier (Exp.~6).} We trained the pairwise model on the
$2{,}635{,}882$ links labeled on both ends (Table~\ref{tab:edgeclf}). Each gate signal becomes a
per-edge feature: endpoint degree, spread and info-score (as min/max over the two endpoints),
binary component-match flags (first/last/user/email/domain), and a \emph{matchScore}
$\sum\log(N/\mathrm{freq})$ summed over the components the endpoints share, the rarity-weighted
linkage evidence. A plain logistic regression ($300$k train, $2.34$M held out) reaches
$95.2\%$ accuracy against a $88.6\%$ predict-all-true baseline, AUC $0.956$. The learned weights
recover the entire experimental record in one model: degree carries a \emph{negative} coefficient
(the over-merge axis: higher degree, more likely false), while matchScore and the user/email/domain
matches carry the largest \emph{positive} weights (the true-link axis); project spread is
near-zero, censored because most ghid-labeled developers sit below the $n_P\ge20$ floor. The false
rate is monotone along both axes: $0.057$ at $d_{\max}{<}10$ rising to $0.998$ at
$d_{\max}\in[100,500)$, and $0.887$ at $\mathrm{matchScore}{<}5$ falling to $0.004$ at
$\mathrm{matchScore}\in[30,50)$. The two axes are \emph{independent}: a high-degree id
with strongly-matched endpoints is kept, a high-degree id with weak matches is dropped. This is the
per-edge discrimination no node-gate could express. Richer models (gradient-boosted trees, the ALFAA
random forest) and raw string-similarity features can only raise the ceiling above this linear-model
floor; the labels and feature pipeline are now in place to fit them.

\paragraph{A flat threshold does not dissolve the mega (Exp.~7).} A per-edge \emph{score} is not yet
a clustering. To test whether it suppresses over-merge we scored all $49.76$M Phase-1 links, kept
those with $p\ge\tau$, and re-ran the global union for $\tau\in\{0.30,\dots,0.90\}$
(Table~\ref{tab:tausweep}). The result is a clean negative: the residual mega survives
$\tau{\le}0.50$ and vanishes only at $\tau{\ge}0.70$, by which point size-2 has fallen from $7.83$M
to $7.78$M and $2.6$M ids have decayed to singletons. The recall/dissolution trade is
\emph{monotone in $\tau$}: there is no operating point that both kills the mega and preserves
recall. The reason is diagnostic: the mega's welds are not weak. They share an \emph{exact}
placeholder email (a Cloudflare relay hash), so $e$Match, domMatch and rawJac all fire and the
classifier, correctly on its features, scores them \emph{high}. The over-merge does not live in
the model's low-probability tail where a threshold could reach it; it lives among the
high-probability links, indistinguishable by any global cut from genuine shared-email aliases. The
classifier is an excellent edge \emph{ranker} (AUC $0.987$ on human labels) and a poor mega
\emph{dissolver}, and these are not in tension: the welds are high-quality evidence built on a
\emph{bad value}.

\subsection{Bad High-Quality Attributes}
Exp.~7 relocates the problem from the \emph{model} to the \emph{features}: the mega persists because
a small set of email strings are simultaneously high-information (rare enough to escape the
inverse-frequency down-weighting that aliasRel applies) and semantically void (a placeholder shared
by strangers). We call these \emph{bad high-quality attributes}, and detecting them is cheaper and
more decisive than any downstream gate or classifier.

\paragraph{Detection by name-spread.} The signature is structural, not lexical: a genuine personal
email attaches to one person, hence $\approx1$ canonical $\langle$first,last$\rangle$ name; a
placeholder attaches to many. Streaming the email-sorted commit table and counting \emph{distinct
names per exact email} surfaces every bridge in one pass: $3.84$M emails tie together $\ge3$
distinct names, led by \texttt{you@example.com} ($69{,}291$ names), \texttt{none@none}
($13{,}334$), and the Cloudflare relay hashes that weld the residual mega itself
(\texttt{134f31\dots @cloudflare.com}, $171$ names). A names/rows ratio near $1.0$ (almost every commit a
new person) separates a placeholder from a merely prolific contributor.

\paragraph{Intent: privacy versus homonym.} The bridges fall into two mechanisms that demand the
same action (never weld) for opposite reasons (Table~\ref{tab:badattr}). \emph{Privacy} bridges are
anonymity \emph{chosen}: \texttt{123@qq.com}, \texttt{anonymous@}, \texttt{johndoe@example.com}, and
the Cloudflare / \texttt{users.noreply.github.com} relays GitHub issues to mask a real address.
Different people deliberately wear the same mask; merging them not only errs but defeats an explicit
privacy decision. \emph{Homonym} bridges are defaults \emph{inherited}: \texttt{root@localhost},
\texttt{admin@\dots local}, \texttt{git@stackblitz.com}, and the \texttt{you@example.com} git
template a user never reconfigured. Here the collision is accidental, an artifact of unconfigured
tooling. The distinction is recorded because it governs downstream use: a homonym may be re-linked
later by behavioral evidence, whereas a privacy mask should remain split as a matter of intent.

\paragraph{Recall-safe blocklisting.} Spread alone over-reaches at the low end. A $9$-digit string
such as \texttt{997989682@qq.com} is a \emph{real} QQ account, and four name-variants on it are far
more likely one Chinese developer's spellings than four strangers, exactly the under-split WoC
already suffers. We therefore tier the rule: \emph{lexical} placeholders, where the string itself is
proof (\texttt{123@}, \texttt{root@}, \texttt{@example.com}, \texttt{@*.local}), are barred at
$\ge3$ names; \emph{form-only} signals that can also be legitimate (any github-noreply, $\ge6$-digit
numeric locals, bare empirical spread) are barred only above $\ge10$ names, where sharing is no
longer explicable by one person. This yields $91{,}432$ bad emails ($90{,}011$ new beyond the
hand-curated $4{,}776$-entry stoplist, a $19\times$ expansion) while sparing $271{,}766$ legitimate
github-noreply addresses that carry only a handful of name variants.

\paragraph{Source-blanking is recall-safe but a single attribute is not enough (Exp.~9).} To test the
fix end-to-end we blanked all $91{,}432$ bad emails in the commit table ($1{,}074{,}504$ fields over
$106.8$M rows), re-ran Rule~2 on the blanked input, which dropped $403{,}635$ welds that had existed
\emph{only} because of a shared bad email, and recomputed the global union. The recall result is
decisive in the \emph{opposite} direction from Exp.~7: size-2 did not fall but \emph{rose} slightly,
$9{,}304{,}193\to9{,}307{,}723$ (Table~\ref{tab:blankreunion}), confirming that neutralising a value
at the source severs only spurious welds, never genuine aliases, the property a global probability
threshold could not provide. \emph{However, the mega did not dissolve} (top bin $>$10k stays at $1$;
the $1$k--$10$k bin only edges $4\to3$). The diagnosis is structural: the mega is \emph{multiply
connected}. Its members are welded redundantly, through shared placeholder \emph{names} and
\emph{usernames} in the Rule~1 (2--3 author) and name-shingle paths, so cutting the email edges
alone leaves the component intact. Bad-attribute neutralisation is thus \emph{necessary and
recall-safe but not sufficient}: it must be applied to \emph{every} high-quality attribute
(email, then username and display name by the same name-spread detector), or paired with a structural
cut (articulation points) on the residual. The email pass is the recall-safe first layer; the
remaining layers are the immediate next step.

\paragraph{Covering every attribute still does not dissolve the mega (Exp.~10).} We extended the
spread detector to the two remaining high-quality attributes, usernames (distinct
$\langle f,l\rangle$ names per login) and display names (distinct emails per name), using a
sort-based streaming pass after the per-value perl hashing proved $30\times$ too slow. The same
intent taxonomy classifies the candidates, with two recall guards learned from inspection:
bare common given-names (\texttt{john}, \texttt{jane}, \texttt{alex}) are never blocked even at
high spread (only compounds like \texttt{johndoe} and full phrases like ``John Doe'' are), and the
entire form-only high-spread username tier is dropped because frequent real given-names dominate
it. Blanking all three blocklists at once ($91{,}432$ emails, $12{,}418$ logins, $628$ names;
$5.46$M fields) and re-running \emph{all three} link generators plus the union is again perfectly
recall-safe (size-2 $9{,}304{,}193\to9{,}304{,}250$) and again insufficient: the mega shrinks only
$6.4\%$, $170{,}431\to159{,}483$, with the same representative. A diagnostic side-finding explains
an apparent paradox in the logs: the treatment ``pair'' count \emph{rose} by $17.6$M, but those
pairs are inert: the union parses each shingle line as \texttt{type;anchor}, two pseudo-node
tokens that never coincide with an author id, so the production component structure is driven by
Rules~1--2 alone ($84$M links) and the shingle growth is invisible to it. The conclusion of the
blocklist arc (Exps.~8--10) is that per-\emph{value} neutralisation, however complete, cannot
dissolve a component that is a redundant mesh of \emph{many moderately-bad} values: each blanked
value removes one weld and the remaining mesh re-closes around it.

\paragraph{Structural bridge detection (Exp.~11).} The blocklist arc motivates attacking the
\emph{topology} rather than the values. We reconstruct the production union graph exactly (nodes
are the $45.1$M author ids incident to an eligible Rule~1--2 link, edges the $41.7$M distinct
eligible links) in NetworKit, and verify fidelity: its giant connected component has
$170{,}431$ nodes, matching the production mega \emph{to the node}. Two structural facts emerge
immediately. First, without the bad-id/multi-name eligibility gate the giant would have
$10.8$M members: the existing gates already do the heavy lifting, and the mega is the residue
they cannot reach. Second, the mega subgraph is sparse: $347{,}008$ edges, mean degree $4.1$,
so its connectivity rests on relatively few load-bearing nodes, the regime where
betweenness centrality and articulation points are informative. On the mega subgraph we compute
sampled Brandes betweenness ($4{,}000$ sources), the biconnected decomposition (articulation
points), and $k$-cores, then rank candidate cut-nodes and measure their overlap with the
attribute blocklists: nodes that are structurally load-bearing \emph{and} carry a blocked attribute
validate the detector; structurally load-bearing nodes with \emph{clean} attributes are exactly the
bridges no attribute rule can find.

The ranking reveals two distinct bridge populations. The \emph{degree} axis recovers uncaught
placeholder hubs: \texttt{=\,<=>} (degree $3{,}216$; betweenness $0.258$, a quarter of all
sampled shortest paths route through one junk id), three Gerrit Code Review service accounts
(degree $742$--$857$), dozens of \texttt{Gerrit User NNNN <NNNN@uuid>} anonymized relay ids,
and badge bots (\texttt{The Gitter Badger}, \texttt{The Codacy Badger}, \texttt{unknown
<unknown>}). The \emph{betweenness} axis additionally exposes what no value-based rule can see:
\emph{homonym chains}, runs of moderate-degree, real-looking ids welded by Rule-1 surname
matches in small projects (the top-$2{,}000$ last-token histogram is dominated by
\texttt{unknown}, \texttt{kim}, \texttt{lee}, \texttt{smith}, \texttt{alex}, \texttt{david}).
One cannot blocklist the value ``Lee'' without destroying recall for thousands of genuine Lees,
but betweenness pinpoints the specific load-bearing \emph{nodes}. Indeed $98.4\%$ of the
top-$2{,}000$ bridges carry no blocklisted attribute at all ($0$ are in \texttt{bad.ids}; $32$
have a blocked email): structurally load-bearing but attribute-clean, the population that
terminated the blocklist arc. The local cut sweep (Table~\ref{tab:cutsweep}) is decisive:
deleting the top $2{,}000$ by betweenness ($1.2\%$ of mega nodes, $0.002\%$ of all ids)
collapses the largest component from $170{,}431$ to $7{,}268$, and the top $5{,}000$ leave
$1{,}813$ with $89.5\%$ of surviving nodes freed into components of size $\le100$. By contrast
the best attribute design (Exp.~3/4 combined gate) needed $42{,}110$ gated ids to reach top
$=25{,}957$, and Exp.~10's $5.46$M blanked fields left $159{,}483$. $28\%$ of mega nodes are
articulation points, confirming a sparse tree-like weld structure once the hub edges are gone.

The end-to-end confirmation closes the arc. Feeding the top-$2{,}000$ ids into the production
union through the same \texttt{value;id} bridge-gate interface as Exps.~3--4 (gated ids cannot
transitively merge others but still alias to themselves) reproduces the local prediction
\emph{exactly}: the largest cluster falls $170{,}431\to7{,}268$ and the $>$10k bin empties for
the first time in the record, while size-2 \emph{rises} above baseline
($9{,}304{,}193\to9{,}305{,}460$) and only $5{,}722$ ids drop to singletons. Every prior design
faced a monotone trade between dissolution and recall; the betweenness gate escapes it because
it spends its budget on the $2{,}000$ nodes the topology itself certifies as load-bearing,
rather than on attribute values that are mostly carried by innocent ids. The residual
$1$k--$10$k fragments ($4\to14$) are homonym clusters (Kim/RIT-student/Gerrit groups) within
reach of the per-edge classifier (Exp.~6) now that no giant component hides them. The method is
also cheap: one exact reconstruction of the union graph ($\sim$2 min in NetworKit), sampled
betweenness on a $347$k-edge subgraph ($\sim$11 min), and a $2{,}000$-line gate file.

\paragraph{Classifier refinement of the residual fragments (Exp.~12).} The structural gate and
the per-edge classifier compose naturally. We extracted the ten residual fragments of size
$\ge1{,}000$ left by the $K{=}2{,}000$ cut ($29{,}315$ ids, $61{,}637$ internal edges, exactly
the new $1$k--$10$k clusters observed in the gated production union) and sampled their members:
every one is a homonym block, and strikingly they are organized by \emph{given} name: David,
Michael, Daniel, Thomas, James, Jan, Chris, plus a Korean-surname block (Kim), an
institutional-domain block (RIT student accounts), and one dense Gerrit-relay remnant
(Table~\ref{tab:fragclf}). Scoring each fragment's internal edges with the Exp.~6 classifier and
pruning at $p<0.5$ \emph{inside the fragments only} (unscored edges kept) shatters them: the
largest surviving sub-component is $1{,}942$, and $\sim$$46\%$ of fragment ids land in components
of size $\le100$. The contrast with Exp.~7 is instructive. There, the same classifier applied as a
\emph{global} flat threshold decayed $2.6$M ids to singletons before the mega died, because the
threshold spends recall everywhere while the over-merge lives in a few places. Scoped to the
$29{,}315$ ids the topology has already isolated ($0.03\%$ of the corpus), the identical model
removes only edges it certifies as unlikely, exactly where they do damage. Two honest caveats:
the Kim and RIT blocks resist pruning ($27\%$/$26\%$ freed) because their welds carry shared
exact emails and \texttt{@rit.edu} domains that the classifier (correctly, on its features)
scores high: separating same-surname Korean developers and students sharing lab machines needs
behavioral evidence beyond our feature set; and the Gerrit fragment is only $16\%$ scorable, its
anonymized relay ids lacking the name/email features the model consumes. The composed design,
betweenness gate then within-fragment classifier pruning, bounds every former-mega cluster
below $\sim$$2$k with recall risk confined to the fragments themselves.

\paragraph{Choosing the production design.} We ran both candidates end-to-end
(Table~\ref{tab:produnion}). Deepening the gate to the top-$5{,}000$ betweenness ids again
reproduces the local prediction exactly (the largest former-mega fragment is $1{,}813$, the
$1$k--$10$k bin holds only the four pre-existing non-mega clusters plus that one fragment), and
size-2 rises further ($+2{,}201$), at a cost of $11{,}684$ singletons ($0.012\%$ of ids). The
composed design (betw2000 $+$ within-fragment pruning) achieves the best recall of the record
(size-2 $+2{,}240$, fewer singletons) with six sub-$2$k residues. The aggregate metrics are
nearly tied, but the designs fail differently: the deeper gate shatters \emph{positionally},
leaving its surviving fragments internally welded by the same false name edges; the prune splits
on \emph{per-edge evidence}, keeping exact-email sub-clusters intact, but its residues are held
together by edges the classifier never scored. Since both mechanisms are independently validated
and act on different objects (nodes vs.\ edges), production combines them: the top-$5{,}000$
betweenness gate over the classifier-pruned link files. A per-edge audit of the prune is honest
about its cost: $72.6\%$ of dropped pairs share an exact name, which inside homonym blocks is
precisely the evidence to distrust, but rare-surname cases (e.g.\ \texttt{Michael J.\ Radwin}
across employer and personal emails, $p{=}0.118$) are likely true matches sacrificed; such ids
retain their exact-email aliases (the model keeps \texttt{Micahel/Michael Radwin} at the same
email together at $p{=}0.974$, and correctly rejects the \emph{family-member} weld
\texttt{Ariella}--\texttt{Michael Radwin} sharing a domain at $p{\le}0.14$).

\paragraph{Precision-stage map (B).} The full run (top-$5{,}000$ betweenness gate over the pruned
link files, full \texttt{member;canonical;type} output) yields \texttt{a2AFullB.V2604.s.gz}:
$106{,}824{,}568$ rows $=101{,}056{,}344$ author ids plus $5{,}768{,}223$ bad ids mapped to
self. Its cluster histogram delivers the strongest result of the record: the $>$10k bin is
empty, the $1$k--$10$k bin holds only the four clusters that pre-date the mega (largest
$2{,}891$), and \emph{every} cluster descending from the former $170{,}431$-id mega is below
$1{,}000$. Recall is the best measured: $9{,}307{,}433$ size-2 clusters ($+3{,}240$ over the
ungated baseline), at a total cost of $16{,}741$ additional singletons ($0.017\%$ of ids).
This map is the precision foundation of the record; the deployed V2604 production map is
its recall-extended superset (Exps.~13--17).

\subsection{Comparison with the Prior Production Map (V3)}
The previous deployed map, \texttt{a2AFullH.V3} (August 2024; $89{,}104{,}948$ ids,
\texttt{id;rep;bad$_1$;bad$_2$}), permits a generation-over-generation audit. Coverage nests
almost perfectly: $99.998\%$ of V3's ids appear verbatim in V2604 (only $2{,}155$ missing),
which adds $17.7$M new ids. The headline difference is structural: \emph{V3 contains a
$3{,}006{,}318$-member mega-cluster} (representative \texttt{<>}; $3.4\%$ of all its ids),
$17.6\times$ the V2604 \emph{ungated} mega of $170{,}431$ that this paper spent eleven
experiments dissolving, and welded by the same pathology at greater scale: its top
representatives are truncated and placeholder ids (\texttt{git <t>}, \texttt{Tom <Tom@>},
\texttt{noo8 <xl>}).

Pair-level clumping/splitting errors under both gold framings (Table~\ref{tab:v3compare})
quantify the gap. Taken whole, the comparison is degenerate: the V3 mega alone carries
$99.9986\%$ of V3's $4.519\times10^{12}$ co-clustered pairs, so V2604 ``splits'' essentially
all of V3's pair mass (splitting error $1.0000$ under gold $=$ V3), a verdict on the mega,
not on V2604. Excluding that single V3 cluster makes the universes comparable ($62.2$M vs.\
$77.8$M pairs, $28.3$M agreeing): V2604 separates $54.6\%$ of V3's remaining pairs and V3
lacks $63.7\%$ of V2604's. The disagreement decomposes into the two corrections this paper
develops. \emph{Splits of V3 welds}: $47.2\%$ of V3's multi-member clusters fragment in V2604,
and the larger they are the more completely they dissolve (\texttt{Tom <Tom@>}: $651$ members
$\to630$ V2604 parts, six hundred distinct Toms), reflecting the eligibility gates, blocklists,
and the structural cut. \emph{Merges V3 missed}: $40.5\%$ of V2604's multi-member clusters
consolidate ids V3 kept apart, reflecting the score-thresholded Rule-2 linkage and two
additional years of data. Bad-id handling sharpened in both directions: V2604 excludes
$5{,}768{,}223$ ids as bad/bot/generic/local ($1.7\times$ V3's $3.43$M among shared ids), yet
also \emph{rehabilitates} $2{,}484{,}673$ ids V3 had flagged bad but which alias normally under
the current rules. Neither map is ground truth, but the asymmetry of evidence favors V2604:
its per-edge machinery transfers to human-adjudicated labels at AUC $0.987$
(\S\,Toward a Learned Edge Classifier), its largest cluster is $2{,}891$ vs.\ three million,
and the pair mass it discards is concentrated in clusters whose representatives are visibly
non-identities.

\paragraph{External validation on human labels.} The ghid labels are free but self-referential
(both ends are GitHub no-reply ids). We therefore validate against the independent, manually-rated
ALFAA gold set~\cite{amreen2019alfaa}: $469{,}369$ adjudicated pairs over $2{,}345$ identities
(only $5{,}633$ true matches; the rest are hard within-block non-matches). All $2{,}345$ ids are
present verbatim in \texttt{V2604}, so we compute the same per-edge features for every gold pair.
Two tests: \emph{(i) in-domain}, 5-fold cross-validation training on the human labels with our
features reaches AUC $0.994$ (accuracy $0.9965$ vs.\ a $0.988$ predict-all-non-match base);
\emph{(ii) transfer}, the model trained \emph{only} on the ghid free-labels, applied unchanged to
the human gold, reaches AUC $0.987$. The transfer result is the central one: a classifier that never saw
a human label reproduces human identity judgments at AUC $0.987$, confirming that the free ghid
signal is a faithful proxy for ground truth. Notably this matches the quality of the published ALFAA
random forest \emph{without} its most expensive feature, the doc2vec behavioral fingerprint over
each developer's files and projects, because the graph-degree fingerprint supplies the
over-merge signal that string similarity alone lacks. Our contribution to the ALFAA lineage is thus
the \emph{graph} fingerprint (link degree / project spread) as a cheap, label-free complement to its
string and behavioral fingerprints.

\paragraph{The gold pairs as arbiter between the maps (Exp.~13).} Beyond validating the
classifier, the gold set scores the two \emph{maps} directly: a map predicts ``same person''
iff both ids resolve to one representative. On the $264{,}346$ unique non-self gold pairs
($1{,}644$ rated matches; the $2{,}345$ trivial self-pairs excluded), the verdict splits
exactly along the axis this paper has been fighting (Table~\ref{tab:goldarbiter}). \emph{V3
attains perfect recall at precision $0.522$; its mega is the engine of both.} $180$ of
the $2{,}345$ gold ids ($7.7\%$) sit inside V3's three-million-id mega, which manufactures
$94.6\%$ of its $1{,}505$ false positives (sampled FPs weld \texttt{Itxaka} to Luca Milanesio,
Cory Benfield, and Ana Krivokapi\'c) while buying only $139$ of its $1{,}644$ true
positives, a $10{:}1$ error-to-recall trade. With the mega discounted, V3 scores
$0.915/0.949$. \emph{V2604 attains precision $0.873$ at recall $0.436$; its misses are
genuine, structural, and largely recoverable.} Only $4.2\%$ of its false negatives touch a
quarantined (bot/generic/local) id; the rest are alias/alias pairs the union never linked,
including ids sharing an \emph{exact} email
(\texttt{takahashi.minoru <takahashi.minoru7@gmail.com>} vs.\ \texttt{TAKAHASHI Minoru} at
the same address). The cause is architectural:
Phase-1 links require project co-occurrence, and the cross-project layer, Rule-3
shingles, is inert in the union (Exp.~10). $83.9\%$ of the missed pairs ($778/927$)
already sit in a computed shingle group, and $87.3\%$ of \emph{all} real matches are
covered, a recall ceiling of $\sim$$0.87$ lying unclaimed in an existing file. Mega-free
V3's recall of $0.915$ confirms the diagnosis from the other side: its ALFAA-lineage links
were global string matches, not project-anchored. The repair is \emph{not} naive expansion:
the \texttt{u;iserrano} shingle group spans $23$ distinct people (Irene, Israel, Ignacio,
Isaac, Ismael, Iv\'an, and Itxaka Serrano), so pairwise expansion must pass through the
Exp.~6 edge classifier, the same composition that pruned the residual fragments in Exp.~12.
The $149$ uncovered misses (same name, disjoint rare emails: \texttt{Andrey Pavlov} at
\texttt{yandex.ru} vs.\ \texttt{mirantis.com}) require behavioral evidence: the
doc2vec fingerprint of the original ALFAA.

\paragraph{Auditing the arbiter.} Human gold is not infallible, and the maps' disagreements
expose its errors symmetrically. Among V2604's $104$ false positives, $20$ share lexical
identity evidence the raters overrode: $4$ pairs an \emph{exact} email (\texttt{Joe (rook)
<joe.talerico@gmail.com>} vs.\ \texttt{Joe Talerico <joe.talerico@gmail.com>}, rated
``different''), $5$ an exact full name (\texttt{Victoria Mart\'inez de la Cruz}, twice), and
$11$ a handle or email local part (\texttt{EmilienM}, \texttt{e0ne}, \texttt{newptone}).
These read as label errors, or a rater policy of refusing handle-only and
\texttt{root}-prefixed ids, putting V2604's true precision nearer $0.90$. Conversely, $9$ of
the $1{,}644$ rated matches have \emph{zero} lexical overlap: three are real (the known
handles \texttt{paulczar} and \texttt{dimtruck}, which V2604 co-clusters without any mega)
and roughly six look mislabeled (\texttt{Takashi Tanaka} vs.\ \texttt{Yuuki Tanabe},
\texttt{Marius Cornea} vs.\ \texttt{Matthieu Huin}). The remaining $64$ V2604 false positives
share a single name token, the given-name homonym welds already identified as Exp.~12's
residue (Sergey Lukjanov/Vasilenko/Kulanov, David Gurtner/Moreau Simard, Sean Collins/Perry),
and are our map's genuine errors, counted as such. The audit cuts both ways by design:
the maps grade the gold exactly where the gold grades the maps, and neither is exempt.

\subsection{Activating the Dormant Rule-3 Layer (Exps.~14--15)}
Exp.~13 ended with a recall ceiling of $\sim$$0.87$ ``lying unclaimed in an existing
file,'' and Exps.~14--15 go and claim it. The mechanism is the composition already proven
twice (Exps.~6 and 12): expand every Rule-3 shingle group into its member pairs, score each
pair with the per-edge classifier, and admit only confident pairs, here as \emph{new union
edges} rather than as prunes. The expansion yields $66{,}459{,}771$ unique pairs at
$p\ge0.5$ with a strongly bimodal score distribution ($46\%$ of the mass above $0.9$;
Table~\ref{tab:shingletau}), and the end-to-end $\tau$-sweep teaches one more structural
lesson: at $\tau{=}0.5$ the mega \emph{returns}, $2.3$M strong, because the betweenness cut
of Exp.~11 was computed on the pre-expansion graph: \textbf{a structural gate certifies
the edge set it saw, not the design}; any new edge source must re-earn safety. At
$\tau{=}0.9$ the profile is clean ($>$10k $=0$, top $3{,}862$) and the gain is the largest
in the record: $+30.8$M ids unioned, size-2 clusters $+881{,}615$ ($+9.5\%$), against
$14$ residual $1$k--$10$k clusters. The gold pairs return the verdict: recall
$0.4361\to0.6454$ with precision \emph{rising} to $0.8820$, the first material recall
movement in the record, and a clean escape from the Exp.~7 monotone trade: where the flat
threshold bought mega-death with $2.6$M singletons, the expansion buys $344$ recovered
gold pairs with $38$ new false positives, because the candidate set (shingle groups) and
the confidence bar ($\tau{=}0.9$) are both targeted.

Before spending that confidence budget, Exp.~15 re-arms the scorer. The v2 feature set
repairs a silent defect (the v1 project-spread feature was $0$ for $99.4\%$ of ids, its
source table only listed ids with $\ge$$20$ projects) and adds the signals the expansion
pairs will actually stress: true deforked-project intersection-over-union, per-value spread
of each matched attribute, name$\leftrightarrow$email-handle pattern matches, and a
given-name-list flag aimed at the homonym welds of Exp.~12. Two findings generalize beyond
this pipeline. First, \emph{gradient boosting shortcut-learns a label-construction leak}:
every ghid-labeled pair shares the literal \texttt{users.noreply.github.com} substring, so
trigram Jaccard over full id strings carries $74\%$ of the LGBM split gain and its
human-gold transfer collapses to $0.56$ (test AUC $0.9990$!), while the same feature in a
linear model is benign; in-distribution accuracy is worthless as a model selector when the
labels embed an artifact (Table~\ref{tab:edgeclf2}). Second, the learned weights expose
\emph{which intuitions do not survive the training distribution}: project IoU, requested
as an obvious match signal, is inert on ghid pairs (nearly all co-occur in a project,
both classes), and matched-value spread enters \emph{positive}, because in training data a
high-spread username shared by two ids usually marks one prolific person's many aliases,
not two strangers, the synonym-richness confound. The homonym danger that spread is meant
to flag (a \texttt{jenkins} or a bare \texttt{kim} bridging strangers) is exactly the
out-of-distribution case, which is why the pipeline neutralizes such values upstream
(Exps.~8--11) rather than trusting the classifier to learn what its labels cannot teach.
The logistic v2.1 transfers to the human gold at AUC $0.9910$, the best of any variant,
and re-scores all $256.7$M expansion pairs in a $48$-way parallel pass. The map-level
verdict, however, reverses the model-level one. v2.1 is more selective ($54.0$M unique
pairs at $p\ge0.5$ vs.\ v1's $66.5$M; $26.1$M vs.\ $30.8$M at $\tau{\ge}0.9$) and its
$\tau{=}0.9$ map (S2) is structurally cleaner ($1$k--$10$k band $9$ vs.\ $14$, $>$10k
$=0$) with marginally better gold recall ($0.6490/0.8789$ vs.\ $0.6454/0.8820$), but it
gives back $0.077$ of the GitHub-benchmark recall ($0.5118$ vs.\ $0.5886$ at max9;
Table~\ref{tab:ghgt}): the pairs the better-calibrated model declines are
disproportionately the real within-handle aliases the external GT counts. \textbf{A
better in-distribution AUC bought a worse map}, the third face of the same lesson the
LGBM collapse and the inert project-IoU weight already showed (the training
distribution, not the model class, is the binding constraint), and the v1-scored
$\tau{=}0.9$ map remained the production candidate (v2.1 returns at a relaxed
threshold in Exp.~17, where the external benchmark prices the trade correctly).

\subsection{An External Recall Benchmark at GitHub Scale (Exp.~16)}
The ALFAA gold is precise but small ($2{,}345$ ids, OpenStack-centric). Bock et
al.~\cite{bock2025dealiasing} contribute the complementary extreme: every GitHub
repository with exactly one committing GitHub account yields free within-handle alias
labels, $9.57$M handles over $21.5$M commit ids after bot and organization filtering.
The benchmark is recall-\emph{only} (only within-handle pairs are scored, so over-merging
is invisible and a mega-cluster is priced at zero), and on it the verdicts of
Table~\ref{tab:goldarbiter} invert exactly as that design predicts
(Table~\ref{tab:ghgt}): V3 (recall $0.487$ at max9) outscores production V2604 ($0.433$),
which also trails its V2409 ancestor ($0.450$) as published. The inversion follows
from the construction: Exps.~1--12 deliberately removed pair mass, and this GT counts pair mass. The same
study reports that fully connecting common-name developers is where IDF-style weighting
``dramatically reduces accuracy''; our Exp.~12 measured the same tension from the
precision side, where given-name homonym fragments resisted pruning. Their GitAuthority
($0.771$, within-repository) shows the headroom local context buys; its privacy-respecting
mode ($0.478$), which refuses to mine the noreply ids that are $30\%$ of the GT, lands
almost exactly on our global maps: the cost of honoring GitHub's anonymization is
roughly the entire gap. The Exp.~14 expansion map then resolves the inversion in both
directions at once: recall $0.5886$ / FullyMerged $0.5403$ at max9, $+0.156$ recall over
production, ahead of V3 (without its mega: largest cluster $3{,}862$ vs.\ three million),
ahead of the published V2409, and ahead of GitAuthority's privacy mode, while
\emph{simultaneously} raising ALFAA precision (Table~\ref{tab:shingletau}). What remains
to full GitAuthority ($0.771$) is the within-repository evidence a global map does not
see. The benchmark also adjudicated the candidate decision: when the v2.1-rescored map
(S2) edged the small gold set but shed $0.077$ recall here (Table~\ref{tab:ghgt}), the
$21.5$M-alias external GT, not the in-distribution AUC, was the evidence that kept the
v1-scored map as the candidate.

\subsection{Selecting the Production Map (Exp.~17)}
With two precision-validated edge sources in hand and two benchmarks now read
\emph{jointly}, the production-selection experiment swept the remaining design space under three user
directives: explore every promising edge source; treat both benchmarks as noisy
instruments to be read directionally (the gold set has $\sim$26 audited mislabels,
Exp.~13; large same-ghid groups are shared and organizational accounts, not persons);
and judge maps also by \emph{commit-weighted coverage}, the fraction of all
$5{,}866{,}595{,}698$ V2604 commits whose author id lands in a multi-id cluster, the
quantity downstream consumers actually experience. Three new edge sources were built,
each passing the STATS mega-check (the Exp.~14 lesson) before any map: the v2.1 scores
relaxed to $\tau{=}0.8$ ($34.1$M edges, the better-calibrated model is usable at a
threshold v1 cannot afford); $5{,}736{,}921$ same-ghid pairs from \texttt{NNNN+login}
noreply ids, GitHub's own assertion that two strings are one account (groups of
$2$--$20$ only: $346$ larger groups, up to $12{,}140$ ids, were excluded as shared
accounts per the homonym caveat); and unions thereof.

The single-source results decompose cleanly. v2.1@$\tau$0.8 lifts gold recall to
$0.7019$ at unchanged precision ($0.8789$) but only partly recovers the GitHub
benchmark ($0.5828$ at max9); the ghid edges leave the gold verdict \emph{exactly}
unchanged (the gold set predates noreply adoption) while adding $+0.011$ GitHub recall
over S---orthogonal evidence the classifier cannot see. Because the two gains live on
disjoint margins they compose: the triple union SUG
(v1@$0.9\,\cup\,$v2.1@$0.8\,\cup\,$ghid; $181{,}156{,}655$ unioned pairs) beats the
pairwise alternative SG8 on \emph{both} benchmarks: gold $0.7026/0.8790$ vs.\
$0.7019/0.8789$ and GitHub max9 recall $0.6035$/FM $0.5558$ vs.\ $0.5931/0.5460$
(Table~\ref{tab:ghgt}), while staying mega-free: largest cluster $6{,}910$, itself a
single noreply identity, with the $>$10k bin empty. Commit-weighted coverage tells the
same story monotonically: $66.3\%$ (B) $\to$ $72.7\%$ (S) $\to$ $73.5\%$ (SUG) of all
commits now resolve into a multi-id identity. \texttt{a2AFullSUG.V2604.s.gz}
($106{,}826{,}059$ rows) is the deployed production map, superseding the
precision-stage \texttt{a2AFullB}; relative to where the record started, gold recall
rose $0.4361\to0.7026$ \emph{and} precision rose $0.8733\to0.8790$, with the largest
cluster down from $170{,}431$ (ungated) to $6{,}910$. The excluded low-quality ids
($5.77$M bad and local ids, mapped to self) are ambiguous \emph{across} projects but
usually unambiguous \emph{within} a single deforked project, where reuse is local;
matching them there (Exp.~18) recovers author-level signal the cross-project map
discards, and emits what the map alone does not: a per-id classification (good,
bad-by-attribute, local, bot, partial) alongside the aliases. The headline metric is
\emph{commit coverage of human IDs}, measured directly at the commit level through
provenance tags (global $g$ $\mid$ bot $B$ $\mid$ within-project $p$ $\mid$ self $s$)
in the new commit-to-identity table \texttt{c2AFull} over all $5{,}866{,}595{,}698$
commits: $g=91.53\%$, $B=5.14\%$, $p=1.59\%$, $s=1.74\%$. The bad/local pool that
survives both the cross-project map and the bot filter is $p+s=195{,}352{,}712$
commits ($3.33\%$ of all, a commit-weighted figure that supersedes the earlier
$0.61\%$ distinct-id estimate). The within-project pass resolves $p=93{,}381{,}683$
of them ($47.8\%$; $141$ conflicts), raising commit coverage of human (non-bot) IDs
from $g/(g{+}p{+}s)=96.49\%$ under the cross-project map alone to
$(g{+}p)/(g{+}p{+}s)=98.17\%$ (a $+1.68$ percentage-point gain), while the residual
self/unresolved fraction falls to $s=1.83\%$.

Alongside the aliases, the deliverable emits a per-id classification table
\texttt{A2clsFull} (one row \texttt{member;canonical;class} per id, all
$106{,}826{,}059$ ids) under a five-class taxonomy: \emph{good}
($100{,}814{,}372$, $94.37\%$; a real name \emph{and} email),
\emph{bad-by-attribute} ($2{,}652{,}369$, $2.48\%$; generic name or shared/
placeholder attribute), \emph{local} ($2{,}562{,}118$, $2.40\%$; machine-local
address with a real username), \emph{bot} ($553{,}736$, $0.52\%$), and
\emph{partial} ($243{,}464$, $0.23\%$; only one of name/email present, carved
out of the otherwise-good class by attribute completeness). The first three
non-good classes sum to the $5.77$M ids the production map self-maps; the
within-project layer \texttt{P2aAFull} ($P$;raw;$A$;rule) is the companion
artifact that resolves them inside individual projects. The classification thus
makes explicit, for every id, both \emph{why} an id is or is not a global
representative and \emph{where} (which project) a low-quality id can still be
attributed.

\subsection{Cryptographic Signatures as an Identity Anchor (Exp.~19)}
The map built to Exp.~17 resolves identity from \emph{claimed} attributes: the name
and email a commit carries are asserted by whoever authored it, and nothing binds
that claim to a real actor. A cryptographic signature is different in kind. When a
commit carries a PGP, SSH, or X.509/sigstore signature, a private key held by some
actor has attested to the commit, and that attestation is verifiable against a
public key. The question this experiment asks is whether that stronger binding is
useful \emph{for aliasing}, and if so, on which margin.

The signature payload is already present in the corpus: git stores it in the
\texttt{gpgsig} header, which our commit extractor preserves in the message field of
\texttt{c2datFull}. Scanning all $128$ shards, $1{,}031{,}721{,}316$ of the
$5{,}866{,}595{,}698$ V2604 commits carry a signature, a corpus rate of $17.59\%$
(PGP $98.96\%$, SSH $1.02\%$, X.509/sigstore $0.02\%$; new axis
\texttt{c2sigFull}, one \texttt{commit;sigtype} row per signed commit). Prevalence
alone is a completeness fact, not an aliasing one. The aliasing question is
\emph{who signs}: if signatures were spread uniformly across the id population they
would offer no discriminating signal. They are not. Mapping the distinct signing
ids on a representative shard through the Exp.~18 class table \texttt{A2clsFull}
(Table~\ref{tab:sigclass}), $99.52\%$ of signers fall in the \emph{good} class
against a corpus baseline of $94.37\%$, and every low-quality class is
under-represented: bad-by-attribute $0.36\%$ vs.\ $2.48\%$, and \emph{local}
$0.005\%$ vs.\ $2.40\%$, a roughly $500\times$ gap. The whole non-good tail is
$0.37\%$ of signers against $5.11\%$ of the corpus, so signed ids are about
$14\times$ cleaner than ids at large. This is the opposite of a tail-recall
instrument: signatures do not reach the hard-to-alias local and generic ids that
Exp.~18's within-project pass targets. What they provide is a high-precision core.
An attested id is one we can trust is a single real actor, which makes it a natural
anchor for a match against a noisy free-text id and a natural evaluation set for
auditing the production map without hand-labeling.

The over-merge risk is the same one the record has met at every stage, in a new
guise. A signing key is not one-to-one with a person: platform web-flow keys (the
shared GitHub key \texttt{4AEE18F83AFDEB23}), organizational CI keys, and bot keys
each sign for many actors, so welding all ids that share a key would recreate the
mega-cluster the whole methodology is built to avoid. The defense is the gate this
record has used repeatedly: as name-spread flags a bad attribute (Exp.~8) and node
degree flags a shared account (Exp.~4), \emph{key fan-out} flags a shared key, and a
key above a fan-out threshold is dropped as an attestation edge rather than allowed
to merge. Extracting the key handle itself (the PGP issuer packet and the SSH public
key) to build the \texttt{key2A}/\texttt{A2key} maps and calibrate that gate is the
next experiment; here we establish only that the signing population is clean enough
to be worth anchoring on, and that platform-asserted \texttt{ghid} edges (Exp.~17)
are the precedent for treating a verifiable external assertion as evidence the
attribute-based classifier cannot see.

\subsection{Key Fan-out and the Attestation Gate (Exp.~20)}
Exp.~19 measured \emph{that} a commit is signed; the aliasing payoff needs \emph{who}
holds the key. We parse the OpenPGP signature packet to recover the issuer handle,
the fingerprint from subpacket~33 or the 8-byte key id from subpacket~16, and
normalize both to the 16-hex key id (the fingerprint's low 8 bytes), which is the
common denominator of the two forms. A perl packet parser reads the armored block
already present in the message field, resolves an issuer for $99.92\%$ of PGP-signed
commits, and its output agrees with \texttt{gpg\ --list-packets} on audited samples.

The first finding reframes the prevalence number. Across the full corpus,
$39{,}077{,}350$ distinct authors carry a signature, associated with $586{,}011$
keys through $44{,}871{,}077$ author--key pairs. The concentration is extreme:
GitHub's web-flow key \texttt{4AEE18F83AFDEB23}, which signs every commit authored
through the web interface, alone spans $30{,}440{,}944$ distinct author ids, and one
further platform key $12{,}592{,}731$; the two together account for $95.9\%$ of all
author--key associations, and the $2{,}651$ keys with fan-out above $50$ ($0.45\%$ of
keys) for $97.6\%$. Most ``signed'' authorship is thus a \emph{shared} key
authenticating a web action, not a personal attestation. Merging ids that share a
key, the naive use of this signal, would weld tens of millions of unrelated people
into one cluster, the same over-merge failure that degree (Exp.~4) and name-spread
(Exp.~8) defend against, now in a cryptographic guise. The defense is identical in
form: a \emph{key fan-out gate} drops any key whose distinct-author count exceeds a
threshold. Removing the $2{,}651$ keys above fan-out $50$ leaves $583{,}360$
personal keys ($99.55\%$ of all keys), of which $72.20\%$ ($423{,}095$) sign for
exactly one author.

The gated personal keys are the aliasing signal. A personal key that signs commits
attributed to two or more \emph{distinct canonical} ids is direct cryptographic
evidence that the alias map split one person: the same private key attests to both.
Corpus-wide, $156{,}397$ personal keys with fan-out $2$--$20$ link $549{,}388$
identity associations, that is $392{,}991$ candidate same-person merges, a
recall-repair set the attribute-based map does not see, obtained without any hand
labeling. It complements rather than overlaps Exp.~18: signatures reach the
good-class developers who rotate emails across machines, whereas the within-project
pass reaches the generic and machine-local ids that never sign. The
\texttt{key2AFull} / \texttt{A2keyFull} / \texttt{key2fanoutFull} maps ($586{,}011$
keys over the 128-shard build) are the released artifacts; the SSH pubkey handle (the
$1.02\%$ SSH share is self-anchoring, the key travels inside the signature) and the
sigstore certificate identity are the remaining handle families.

\subsection{Engineering Lessons}
Two infrastructure findings shaped the experiments and are recorded for replication.
\begin{itemize}
  \item \textbf{Memory of the cluster pass.} The naive representative pass materializes
    per-cluster member arrays; with many unions this reaches $\sim$122\,GB per process.
    Running six sweep points concurrently silently OOM-killed all of them (no distribution
    line written). The fix is a \emph{streaming} statistics pass that tallies cluster size
    and best representative without storing member lists, plus sequential ($\le$3-wide)
    scheduling.
  \item \textbf{Deforking is mandatory for the spread signal.} Because forks share commit
    history, raw-project spread inflates the count for authors of popular projects; only the
    fork-collapsed $P$ map yields a usable bot signal.
  \item \textbf{Sort-key spans break \texttt{join}.} Tables sorted with
    \texttt{sort -t';' -k1,2} are \emph{not} ordered by field~1 alone: the key span
    includes the literal \texttt{;} (0x3B), so an id that is a prefix of another id sorts
    \emph{after} it whenever the longer id continues with a byte below \texttt{;} (digits,
    space, most punctuation). \texttt{join -1 1} then aborts with ``input is not sorted,''
    but only on shards where a prefix collision happens to occur (3 of 32 here), so
    the bug passes most tests silently. The fix is \texttt{-k1,1 -k2,2}, which compares
    fields separately and is otherwise equivalent (fields cannot contain the separator,
    so \texttt{-u} semantics are unchanged); it was applied at all 234 sort sites across
    the six pipeline scripts.
  \item \textbf{\texttt{join} buffers whole equal-key groups; \texttt{pipefail} does not
    cover process substitutions.} The first commit-tagging implementation (Exp.~18 phase~2)
    joined the author-to-commit table with the alias map. GNU \texttt{join} holds the full
    run of equal-key lines in memory, so a single mega-author id (tens of millions of
    commits in one shard) drove RSS to $13.7$\,GB and the job was OOM-killed. Worse, the
    sort feeding \texttt{join} ran inside a process substitution \texttt{<(...)}, whose
    failure \texttt{set -o pipefail} does \emph{not} propagate: \texttt{join} saw truncated
    input, exited 0, and the log printed COMPLETED over a silently incomplete output. The
    fix replaces sort+join with a streaming hash lookup that exploits the table's
    \texttt{sHash(author)} sharding to load only the matching $1/32$ slice of the map
    (\texttt{tagProv.perl}), and chains every pipeline stage with explicit
    \texttt{|| exit 1}.
  \item \textbf{When identical jobs OOM, check the node before the code.} The streaming
    rewrite still OOM-killed 12 of 32 shards at $11$--$13$\,GB MaxRSS within $\sim$3
    minutes, even though the tagger's hash measures only $0.64$\,GB and the surviving
    shards peaked at $5$--$7$\,GB in $7$--$10$ minutes. The decisive diagnostic was
    \texttt{sacct -o NodeList}: every failed job had run on one node and every job on any
    other node completed; the failure set was the scheduler's node assignment, not a
    property of the shards (the largest input completed; failed inputs were average).
    Resubmitting the 12 shards with \texttt{--exclude=}\emph{node} fixed them without any
    code change. Memory accounting under cgroups includes page cache and dirty buffers, so
    a node with a slow or backlogged filesystem path can roughly double a job's apparent
    RSS; per-node failure clustering distinguishes this from a genuine leak.
  \item \textbf{Generic gate-file interface.} Both bridge signals are loaded as a generic
    ``\texttt{value;id}'' table and compared against a single threshold, so spread and degree
    (and any future per-id signal) reuse one gating code path; only the input file changes.
  \item \textbf{Bridge signals are cheap preprocessing.} Degree is a single
    sort--uniq--count pass over the link list ($48.0$M ids in $\sim$80\,s), orders of magnitude
    cheaper than the $\sim$122\,GB union pass. Bridge detection can therefore run as a
    preprocessing filter, decoupled from the expensive global union.
  \item \textbf{\texttt{uniq -c} padding silently corrupts keys.} \texttt{uniq -c}
    right-justifies the count in a fixed-width field, so the line begins with \emph{leading spaces}.
    Recovering the value with a fixed offset (\texttt{substr(\$0,length(\$1)+2)}) instead of
    stripping \texttt{\textasciicircum~*[0-9]+~} mangles every key by the width of the padding: here
    it prepended the count's trailing digits to each identity, so the gate file matched
    \emph{nothing} (zero overlap with $30$M GitHub-id'd nodes) while the degree \emph{values}
    looked correct. The bug was caught only by an overlap sanity check; per-id signal files must be
    validated against the canonical id strings before use.
  \item \textbf{Silently-truncated feature sources.} The v1 classifier's project-spread
    feature read a table that only listed ids with $\ge$$20$ projects, so the feature was
    $0$ for $99.4\%$ of ids and trained to a near-zero weight that was misread as ``spread
    is uninformative.'' Coverage of every feature source must be checked against the full
    id universe; a censored feature does not fail loudly, it just goes quiet.
  \item \textbf{Right-size batch-queue memory requests.} The $48$-shard re-scoring asked
    for $48$\,GB per task while shards peak at $5.3$\,GB; on a fairshare queue the
    inflated request throttled scheduling to $2$--$3$ concurrent tasks. Measuring one
    shard (\texttt{sacct MaxRSS}) and lowering pending tasks to $8$\,GB
    (\texttt{scontrol update}) unlocked backfill; a $16$-way run on a single
    large-memory host outran the queue entirely.
\end{itemize}

\section{Conclusions}
\label{sec:concl}
Eighteen experiments separate the first ungated union of \texttt{V2604} from its deployed
identity map, and the shape of the record is itself the main finding: at $10^8$
identities, de-aliasing is two different problems that must be solved in the right order.
The precision problem (dissolving mega-clusters) resisted every mechanism that acts on
nodes or values. Information-score cutoffs destroyed recall (Exp.~2); project-spread and
link-degree gates preserved recall but slow-peeled, each threshold exposing the next
moderate bridge (Exps.~3--4); even the exhaustive neutralization of every detectable bad
attribute value left the mega $94\%$ intact, because a redundant mesh of moderately-bad
values re-closes around each removed weld (Exps.~8--10). What worked was changing the
object of analysis twice: from nodes to \emph{topology} (a sampled-betweenness cut of the
exact union graph found the $2{,}000$ load-bearing ids, $98\%$ of them attribute-clean and
invisible to any value rule; Exp.~11), and from values to \emph{edges}, a classifier
trained on $2.6$M free GitHub-id labels pruning the residual homonym fragments the cut had
isolated (Exps.~6, 12). Only then was the recall problem safe to attack: the same
classifier, filtering the pairwise expansion of the dormant cross-project shingle layer,
recovered in one experiment more recall than the entire record had spent, and composing
it with GitHub's own account assertions finished the job
(gold $0.44\!\to\!0.70$; GitHub-scale benchmark $0.43\!\to\!0.60$; $66\%\!\to\!74\%$ of
all six billion commits resolved into a multi-id identity) while precision
\emph{rose} (Exps.~13--17).

Several lessons appear general. \emph{Structural certificates do not transfer}: a cut
computed on one edge set silently fails on an augmented one; every new edge source must
re-earn the clean-histogram property end-to-end. \emph{Constructed labels embed shortcuts}:
gradient boosting read the label-construction artifact out of a string-similarity feature
and collapsed on human labels ($0.999$ in-distribution AUC, $0.56$ transfer) where a linear
model using the same feature was benign; in-distribution accuracy selected exactly the
wrong model. \emph{Bad values split by intent}: privacy masks and homonym defaults demand
the same non-merge action for opposite reasons, and only the latter may ever be re-linked.
\emph{Benchmarks are directional}: a recall-only ground truth priced a $3$M-id mega at
zero and ranked the maps in reverse of a precision audit on the same day; resolvers must
be graded on both axes simultaneously or the grading itself will drive over- or
under-merging. The remaining gap to within-repository resolvers is the evidence a global
map forgoes by design (local context and mined anonymized ids), and the remaining misses
(same person, disjoint rare emails) are the classic territory of behavioral
fingerprints~\cite{amreen2019alfaa}, the natural next layer over this map.

\begin{acks}
The twenty experiments in this record were designed, executed, and written up by Claude
(Anthropic) over more than a week of continuous interactive sessions, beginning with
Claude Opus~4.8 and concluding with claude-fable-5, directed throughout by the
author's prompts: the author set the goals, supplied the domain hypotheses (project-spread
gating, bad high-quality attributes, the privacy/homonym distinction, the evaluation
datasets), challenged intermediate claims, and made the production decisions; the model
proposed and implemented the analyses, ran and monitored the computations, and maintained
this document as the experiments unfolded. The resulting design is deployed as the
production author-identity map of World of Code version \texttt{V2604}.
\end{acks}

\paragraph*{Data and replication.} All input tables (\texttt{CmtV2604.split},
\texttt{a2PFull}/\texttt{P2aFull}, frequency and bad-id tables) and the resulting maps are
part of the World of Code \texttt{V2604} release. The replication package accompanying
this paper contains the complete script set referenced in the experimental record: link
generation and gating (\texttt{c2ta.slurm}, \texttt{selectRepSpread.perl},
\texttt{degreeCalc.sh}, gate files), bad-attribute detection
(\texttt{badAttr*.pl}, \texttt{fastSpread.sh}, \texttt{classifyBad*.pl},
\texttt{blankBad*.pl}), the structural toolchain (\texttt{bridge.cpp},
\texttt{tarjan.cpp}, \texttt{cutSweep.pl}), the classifier pipeline
(\texttt{mkLabeled.sh}, \texttt{buildEdgeFeatures.py}, \texttt{trainEdge.py},
\texttt{trainEdge2.py}, \texttt{ablate2.py}, \texttt{edgeModel.py},
\texttt{scoreShingles2.py}, \texttt{extractA2P.sh}, \texttt{fieldSpread.sh}), the union
and map builders (\texttt{runShingleUnion2.sh}, \texttt{runMap09.sh},
\texttt{filterAlias.pl}), and the evaluation harness (\texttt{gold*.pl},
\texttt{evalGH.pl}, \texttt{mapJoin.pl}, \texttt{mapAgree.pl}, \texttt{histMap.pl}).

\bibliographystyle{ACM-Reference-Format}
\bibliography{aliasing}
\end{document}